\documentclass[5p]{elsarticle}

\journal{X}

\usepackage{amsmath,amssymb,amsfonts}
\usepackage{algorithmic}
\usepackage{graphicx}
\usepackage{tikz}
\usepackage{textcomp}
\usepackage{xcolor}
\usepackage{subcaption}
\usepackage{booktabs}
\usepackage{tabularx}
\usepackage{siunitx}
\usepackage{acro}[=v2]
\usepackage{colortbl}
\usepackage{listings}
\usepackage{newtxmath}
\usepackage{csquotes}
\usepackage{url}
\DeclareAcronym{acsrp}{
	short	= AC ,
	long	= admission control ,
}

\DeclareAcronym{af}{
	short	= AF ,
	long	= assured forwarding ,
}

\DeclareAcronym{amf}{
	short	= AMF ,
	long	= access and mobility management function ,
}

\DeclareAcronym{api}{
	short 	= API ,
	long	= application programming interface ,
}

\DeclareAcronym{asic}{
	short	= ASIC ,
	long	= application-specific integrated circuit ,
}

\DeclareAcronym{atcll}{
	short	= ATCLL ,
	long	= Aveiro Tech City Living Lab ,
}

\DeclareAcronym{ats}{
    short   = ATS ,
    long    = asynchronous traffic shaper ,
}

\DeclareAcronym{av}{
    short   = AV ,
    long    = audio and video ,
}

\DeclareAcronym{avb}{
    short   = AVB ,
    long    = audio/video bridging ,
}

\DeclareAcronym{be}{
	short	= BE ,
	long	= best-effort ,
}

\DeclareAcronym{bmc}{
    short   = BMC ,
    long    = best master clock ,
}

\DeclareAcronym{bod}{
	short	= BoD ,
	long	= bandwidth on demand ,
}

\DeclareAcronym{cap}{
	short	= CAP ,
	long	= common alerting protocol ,
}

\DeclareAcronym{cbe}{
	short	= CBE ,
	long	= cell broadcast entity ,
}

\DeclareAcronym{cbc}{
	short	= CBC ,
	long	= cell broadcast center ,
}

\DeclareAcronym{cbr}{
	short	= CBR ,
	long	= constant bit rate ,
}

\DeclareAcronym{cbservice}{
	short	= CBS ,
	long	= cellular broadcast service ,
}

\DeclareAcronym{cbs}{
    short   = CBS ,
    long    = credit-based shaper ,
}

\DeclareAcronym{ci}{
    short   = CI ,
    long    = congestion isolation ,
}

\DeclareAcronym{cmas}{
	short	= CMAS ,
	long	= Commercial Mobile Alert System ,
}

\DeclareAcronym{cnc}{
	short	= CNC ,
	long 	= centralized network configuration ,
}

\DeclareAcronym{cuc}{
	short	= CUC ,
	long	= centralized user configuration ,
}

\DeclareAcronym{cots}{
    short   = COTS ,
    long    = commercial off-the-shelf ,
}

\DeclareAcronym{cpu}{
	short	= CPU ,
	long	= central processing unit ,
}

\DeclareAcronym{cqf}{
    short   = CQF ,
    long    = cyclic queueing and forwarding ,
}

\DeclareAcronym{dcu}{
	short	= DCU ,
	long	= data collecting unit ,
}

\DeclareAcronym{ds-tt}{
    short   = DS-TT ,
    long    = device-side TSN translator ,
}

\DeclareAcronym{ea}{
	short	= EA ,
	long	= emergency area ,
}

\DeclareAcronym{eas}{
	short	= EAS ,
	long	= emergency alerting system ,
}

\DeclareAcronym{eena}{
	short	= EENA ,
	long	= European Emergency Number Association ,
}

\DeclareAcronym{ef}{
	short	= EF ,
	long	= expedited forwarding ,
}

\DeclareAcronym{enb}{
	short	= eNB ,
	long	= eNodeB ,
}

\DeclareAcronym{etsi}{
	short	= ETSI ,
	long 	= European Telecommunications Standards Institute ,
}

\DeclareAcronym{etws}{
	short	= ETWS ,
	long	= Earthquake and Tsunami Warning System ,
}

\DeclareAcronym{fp}{
    short   = FP ,
    long    = frame preemption ,
}

\DeclareAcronym{fpga}{
	short	= FPGA ,
	long	= fully programmable gate array ,
}

\DeclareAcronym{frer}{
    short   = FRER ,
    long    = frame replication and elimination for reliability ,
}

\DeclareAcronym{gb}{
    short   = GB ,
    long    = guard band ,
}

\DeclareAcronym{gcl}{
    short   = GCL ,
    long    = gate control list ,
}

\DeclareAcronym{gm}{
    short   = GM ,
    long    = grand-master ,
}

\DeclareAcronym{gnb}{
	short	= gNB ,
	long	= gNodeB ,
}

\DeclareAcronym{gps}{
	short	= GPS ,
	long	= global positioning system ,
}

\DeclareAcronym{gptp}{
    short   = gPTP ,
    long    = generalized precision time protocol ,
}

\DeclareAcronym{hsr}{
    short   = HSR ,
    long    = high-availability seamless redundancy ,
}

\DeclareAcronym{ieee}{
	short	= IEEE ,
	long	= Institute of Electrical and Electronics Engineers ,
}

\DeclareAcronym{ilp}{
    short   = ILP ,
    long    = integer linear programming ,
}

\DeclareAcronym{io}{
	short				= I/O ,
	long				= input and output ,
	short-plural-form	= I/Os ,
	long-plural-form 	= inputs and outputs ,
}

\DeclareAcronym{iot}{
	short	= IoT ,
	long	= Internet of things ,
}

\DeclareAcronym{ip}{
	short	= IP ,
	long	= internet protocol ,
}

\DeclareAcronym{isis}{
    short   = IS-IS ,
    long    = intermediate station to intermediate station ,
}

\DeclareAcronym{itu}{
	short	= ITU ,
	long	= International Telecommunication Union ,
}

\DeclareAcronym{json}{
	short	= JSON ,
	long	= JavaScript Object Notation ,
}

\DeclareAcronym{lafsrp}{
	short	= LAF ,
	long	= listener asking failed ,
}

\DeclareAcronym{lbsms}{
	short	= LB-SMS ,
	long	= location-based SMS ,
}

\DeclareAcronym{lrsrp}{
	short	= LR ,
	long	= listener ready ,
}

\DeclareAcronym{l2}{
	short	= L2 ,
	long	= layer 2 ,
}

\DeclareAcronym{m2m}{
    short   = M2M ,
    long    = machine-to-machine ,
}

\DeclareAcronym{mlc}{
	short	= MLC ,
	long	= mobile location center ,
}

\DeclareAcronym{mme}{
	short	= MME ,
	long	= mobility management entity ,
}

\DeclareAcronym{mpls}{
	short	= MPLS ,
	long	= multiprotocol label switching ,
}

\DeclareAcronym{nic}{
	short	= NIC ,
	long	= network interface card ,
}

\DeclareAcronym{nist}{
	short	= NIST ,
	long	= National Institute of Standards and Technology ,
}

\DeclareAcronym{nos}{
	short 	= NOS ,
	long	= network operation system ,
}

\DeclareAcronym{nw-tt}{
    short   = NW-TT ,
    long    = network-side TSN translator ,
}

\DeclareAcronym{obu}{
	short	= OBU ,
	long	= on-board unit ,
}

\DeclareAcronym{onf}{
	short	= ONF ,
	long	= Open Networking Foundation ,
}

\DeclareAcronym{opcua}{
    short   = OPC-UA ,
    long    = Open Platform Communications unified architecture ,
}

\DeclareAcronym{os}{
	short	= OS ,
	long	= operative system ,
}

\DeclareAcronym{osi}{
	short 	= OSI ,
	long	= Open Systems Interconnection ,
}

\DeclareAcronym{pbr}{
	short	= PBR ,
	long	= policy-based routing ,
}

\DeclareAcronym{pcr}{
    short   = PCR ,
    long    = path control and reservation ,
}

\DeclareAcronym{pcp}{
    short   = PCP ,
    long    = priority code point ,
}

\DeclareAcronym{pdv}{
	short	= PDV ,
	long	= packet delay variation ,
}

\DeclareAcronym{prp}{
    short   = PRP ,
    long    = parallel redundancy protocol , 
}

\DeclareAcronym{psfp}{
    short   = PSFP ,
    long    = per-stream filtering and policing ,
}

\DeclareAcronym{ptp}{
	short	= PTP ,
	long 	= precision time protocol ,
}

\DeclareAcronym{pws}{
	short	= PWS ,
	long	= public warning system ,
}

\DeclareAcronym{p4}{
	short	= P4 ,
	long	= programming protocol-independent packet processor ,
}

\DeclareAcronym{qos}{
	short	= QoS ,
	long 	= quality-of-service ,
}

\DeclareAcronym{rap}{
    short   = RAP ,
    long    = resource allocation protocol ,
}

\DeclareAcronym{rest}{
	short	= REST ,
	long	= representational state transfer ,
}

\DeclareAcronym{rnc}{
	short	= RNC ,
	long	= 3G radio network controller ,
}

\DeclareAcronym{rsu}{
	short	= RSU ,
	long	= road-side unit ,
}

\DeclareAcronym{rsvp}{
	short	= RSVP ,
	long	= resource reservation protocol, 
}

\DeclareAcronym{rsvpte}{
	short	= RSVP-TE ,
	long	= RSVP traffic engineering ,
}

\DeclareAcronym{rtt}{
	short	= RTT ,
	long	= round-trip time,
}

\DeclareAcronym{seu}{
    short   = SEU ,
    long    = single-event upset ,
}

\DeclareAcronym{sdn}{
	short	= SDN ,
	long	= software-defined network ,
}

\DeclareAcronym{sms}{
	short	= SMS ,
	long	= short message service ,
}

\DeclareAcronym{smt}{
    short   = SMT ,
    long    = satisfiability modulo theory
}

\DeclareAcronym{spr}{
    short   = SPR ,
    long    = shortest path routing ,
}

\DeclareAcronym{srp}{
	short	= SRP ,
	long	= stream reservation protocol ,
}

\DeclareAcronym{ta}{
	short	= TA ,
	long	= tracking area ,
}

\DeclareAcronym{tas}{
    short   = TAS ,
    long    = time-aware shaper ,
}

\DeclareAcronym{tasrp}{
	short	= TA ,
	long	= talker advertise ,
}

\DeclareAcronym{tfsrp}{
	short	= TF ,
	long	= talker failed ,
}

\DeclareAcronym{tcp}{
	short	= TCP ,
	long 	= transmission control protocol ,
}

\DeclareAcronym{tdma}{
	short	= TDMA ,
	long	= time division multiple access ,
}

\DeclareAcronym{tls}{
	short	= TLS ,
	long	= transport layer security ,
}

\DeclareAcronym{tsn}{
	short	= TSN ,
	long	= time-sensitive network ,
}

\DeclareAcronym{tte}{
    short   = TTE ,
    long    = time-triggered Ethernet ,
}

\DeclareAcronym{uav}{
    short   = UAV ,
    long    = unmanned aerial vehicle ,
}

\DeclareAcronym{ubs}{
    short   = UBS ,
    long    = urgency-based scheduler ,
}

\DeclareAcronym{ue}{
	short				= UE ,
	long				= user equipment ,
	long-plural-form 	= user equipment ,
}

\DeclareAcronym{udp}{
	short	= UDP ,
	long	= user datagram protocol ,
}

\DeclareAcronym{ull}{
    short   = ULL ,
    long    = ultra-low latency ,
}

\DeclareAcronym{upf}{
    short   = UPF ,
    long    = user-plane function ,
}

\DeclareAcronym{v2v}{
    short   = V2V ,
    long    = vehicle-to-vehicle ,
}

\DeclareAcronym{vanet}{
	short	= VANET ,
	long	= vehicular ad-hoc network ,
}

\DeclareAcronym{vlan}{
	short	= VLAN ,
	long 	= virtual local area network ,
}

\DeclareAcronym{wan}{
	short	= WAN ,
	long	= wide area network ,
}

\DeclareAcronym{xml}{
	short	= XML ,
	long 	= extensible markup language ,
}

\DeclareAcronym{3gpp}{
	short	= 3GPP , 
	long	= 3rd Generation Partnership Project ,
}

\DeclareAcronym{3g}{
	short	= 3G ,
	long	= third generation ,
}

\DeclareAcronym{4g}{
	short	= 4G ,
	long	= fourth generation ,
}

\DeclareAcronym{5g}{
	short	= 5G ,
	long	= fifth generation ,
}

\usepackage{amsmath}
\usepackage{amssymb}
\usepackage{amsfonts}

\DeclareAcronym{max}{
	short	= \ensuremath{\max\{x\}},
	long 	= maximum \ensuremath{x} value,
	class	= common-symbols
}

\DeclareAcronym{max_alt1}{
	short	= \ensuremath{x_{\max}},
	long 	= maximum \ensuremath{x} value,
	class	= common-symbols
}

\DeclareAcronym{max_alt2}{
	short	= \ensuremath{\hat x},
	long 	= maximum \ensuremath{x} value,
	class	= common-symbols
}

\DeclareAcronym{min}{
	short	= \ensuremath{\min\{x\}},
	long 	= minimum \ensuremath{x} value,
	class	= common-symbols
}

\DeclareAcronym{min_alt1}{
	short	= \ensuremath{x_{\min}},
	long 	= minimum \ensuremath{x} value,
	class	= common-symbols
}

\DeclareAcronym{min_alt2}{
	short	= \ensuremath{\check x},
	long 	= minimum \ensuremath{x} value,
	class	= common-symbols
}

\DeclareAcronym{lenec}{
	short 	= \ensuremath{\lambda_{c}^{e}},
	long 	= length of entity \ensuremath{e} in context \ensuremath{c},
	class	= network-symbols
}

\DeclareAcronym{lene}{
	short 	= \ensuremath{\lambda_{c}},
	long 	= length of entity \ensuremath{e},
	class	= network-symbols
}

\DeclareAcronym{packet}{
	short 	= \ensuremath{p},
	long 	= a packet,
	class	= network-symbols
}

\DeclareAcronym{packets}{
	short 	= \ensuremath{P},
	long 	= a collection of packets,
	class	= network-symbols
}

\DeclareAcronym{spackets}{
	short 	= \ensuremath{P'},
	long 	= a sub-collection of packets,
	class	= network-symbols
}

\DeclareAcronym{sspackets}{
	short 	= \ensuremath{P'},
	long 	= a sub-collection of packets,
	class	= network-symbols
}

\DeclareAcronym{ratio}{		% TODO variable name.
	short	= \ensuremath{r},
	long	= ratio of ,
	class 	= network-symbols
}

\DeclareAcronym{rate}{		% TODO variable name.
	short	= \ensuremath{f},
	long	= rate of ,
	class 	= network-symbols
}

\DeclareAcronym{slope}{		% TODO variable name.
	short	= \ensuremath{s},
	long	= slope of ,
	class 	= network-symbols
}

\DeclareAcronym{period}{		% TODO variable name.
	short	= \ensuremath{T},
	long	= period of ,
	class 	= network-symbols
}

\DeclareAcronym{edges}{
	short	= \ensuremath{E},
	long	= set of edges ,
	class	= graph-symbols
}

\DeclareAcronym{vertices}{
	short	= \ensuremath{V},
	long	= set of vertices ,
	class	= graph-symbols
}

\DeclareAcronym{graphve}{
	short	= \ensuremath{G(V,E)},
	long	= graph where \ensuremath{E \subseteq  \{e\} = \{\{u,v\} \,:\,u,v \in V \text{ and } u \neq v\}} ,
	class	= graph-symbols
}

\DeclareAcronym{psnr}{
	short	= \ensuremath{\mathrm{PSNR}},
	long	= peak signal-to-noise ratio ,
	class	= kpi-symbols
}

\DeclareAcronym{sinr}{
	short	= \ensuremath{\mathrm{SINR}},
	long	= signal-interference-to-noise ratio ,
	class	= kpi-symbols
}

\DeclareAcronym{snr}{
	short	= \ensuremath{\mathrm{SNR}},
	long	= signal-to-noise ratio ,
	class	= kpi-symbols
}

\DeclareAcronym{ssim}{
	short	= \ensuremath{\mathrm{SSIM}},
	long	= structural similarity index measure ,
	class	= kpi-symbols
}

\DeclareAcronym{pred-label}{
	short	= \ensuremath{\hat{\mathbf{y}}},
	long	= label predicted by a function \ensuremath{f},
	class	= ml-symbols
}

\DeclareAcronym{rank_priority}{
	short	= \ensuremath{p},
	long	= priority level,
	class	= rank-symbols
}

\DeclareAcronym{rank_requirements}{
	short	= \ensuremath{R},
	long 	= list of requirements, 
	class	= rank-symbols
}

\DeclareAcronym{rank_requirement_cap}{
	short	= \ensuremath{\rho},
	long 	= requirement assessment, 
	class	= rank-symbols
}

\DeclareAcronym{rank_bid_div_constant}{
	short	= \ensuremath{\tau},
	long	= division constant for requirements ,
	class	= rank-symbols
}

\DeclareAcronym{rank_bid_div_constant_min}{
	short	= \ensuremath{\tau_{\min}},
	long	= minimum division constant for requirements ,
	class	= rank-symbols
}

\DeclareAcronym{rank_bid_div_constant_max}{
	short	= \ensuremath{\tau_{\max}},
	long	= maximum division constant for requirements ,
	class	= rank-symbols
}

\DeclareAcronym{rank_bid_function}{
    short   = \ensuremath{\alpha\left(\symbrank{requirements}\right)} ,
    long    = current node resources assessment function for a list of requirements \symbrank{requirements} ,
    class   = symbols-rank 
}

\DeclareAcronym{rank_bid_function_reduction}{
    short   = \ensuremath{\alpha\left(\symbrank{requirements}_{1..*}\right)} ,
    long    = current node resources assessment function for a smaller list of requirements \ensuremath{\symbrank{requirements}_{1..*}} ,
    class     = symbols-rank 
}

\DeclareAcronym{rank_rtt_function}{
	short	= \ensuremath{\phi_1\left(x, \symbrank{rtt_threshold}\right)} ,
	long	= function of round-trip time assessment of tested value \ensuremath{x} in comparison to the defined \symbrank{rtt_threshold} threshold value ,
	class	= symbols-rank
}

\DeclareAcronym{rank_rtt_threshold}{
	short	= \ensuremath{r} ,
	long	= round-trip time threshold for assessment ,
	class	= symbols-rank
}

\DeclareAcronym{rank_hops_function}{
	short	= \ensuremath{\phi_2\left(x\right)} ,
	long	= function of number of hops assessment of tested value \ensuremath{x} ,
	class	= symbols-rank
}

\DeclareAcronym{rank_jitter_function}{
	short	= \ensuremath{\phi_3\left(x, \symbrank{pdv_threshold}, \symbrank{pdv_tolerance}\right)} ,
	long	= function of jitter assessment of tested value \ensuremath{x} ,
	class	= symbols-rank
}

\DeclareAcronym{rank_packet_loss_function}{
	short	= \ensuremath{\phi_4\left(x, \symbrank{pl_threshold}, \symbrank{pl_m_i}\right)} ,
	long	= function of packet loss assessment of tested value \ensuremath{x} ,
	class	= symbols-rank
}

\DeclareAcronym{rank_pdv_threshold}{
	short	= \ensuremath{j} ,
	long	= maximum admitted packet delay variance threshold value for assessment ,
	class	= symbols-rank
}

\DeclareAcronym{rank_pdv_tolerance}{
	short	= \ensuremath{j^{+}} ,
	long	= tolerance in percentage of the accepting threshold value for assessment ,
	class	= symbols-rank
}

\DeclareAcronym{rank_pl_threshold}{
	short	= \ensuremath{\lambda} ,
	long	= maximum admitted packet loss threshold value for assessment ,
	class	= symbols-rank
}

\DeclareAcronym{rank_pl_variable}{
	short	= \ensuremath{\sigma} ,
	long	= impact rate choice for packet loss assessment ,
	class	= symbols-rank
}

\DeclareAcronym{rank_pl_m_1}{
	short	= \ensuremath{\symbrank{pl_variable}_1} ,
	long	= strong degradation method for packet loss assessment ,
	class	= symbols-rank
}

\DeclareAcronym{rank_pl_m_2}{
	short	= \ensuremath{\symbrank{pl_variable}_2} ,
	long	= linear degradation method for packet loss assessment ,
	class	= symbols-rank
}

\DeclareAcronym{rank_pl_m_3}{
	short	= \ensuremath{\symbrank{pl_variable}_3} ,
	long	= smooth degradation method for packet loss assessment ,
	class	= symbols-rank
}

\DeclareAcronym{rank_salt_valorization}{
	short	= \ensuremath{\vartheta} ,
	long	= salt valorization ,
	class	= symbols-rank
}

\DeclareAcronym{rank_pl_m_i}{
	short	= \ensuremath{\symbrank{pl_variable}_i} ,
	long	= i-degradation method for packet loss assessment ,
	class	= symbols-hidden
}

\DeclareAcronym{rank_bid_history_1}{
	short 	= \ensuremath{a_1} ,
	long	= number of requirements on the average number of requested resource requirements ,
	class 	= symbols-rank
}

\DeclareAcronym{rank_bid_history_2}{
	short 	= \ensuremath{a_2} ,
	long	= percentage of non-used-but-allocated resource reservations ,
	class 	= symbols-rank
}

\DeclareAcronym{rank_bid_history_3}{
	short 	= \ensuremath{a_3} ,
	long	= average duration of last known similar admission request over the average duration of all ,
	class 	= symbols-rank
}

\DeclareAcronym{rank_bid_history_4}{
	short 	= \ensuremath{a_4} ,
	long	= percentage of strict resource reservations over all allowed admissions ,
	class 	= symbols-rank
}

\DeclareAcronym{rank_bid_history_salt}{
	short	= \ensuremath{s} ,
	long 	= salt random value as contributor to \symbrank{bid_performance_history} ,
	class	= symbols-rank
}

\DeclareAcronym{rank_bid_history_weight}{
	short	= \ensuremath{\delta} ,
	long 	= weight of each variable of \symbrank{bid_performance_history} contributors ,
	class	= symbols-rank
}

\DeclareAcronym{rank_bid}{
	short 	= \ensuremath{B} ,
	long	= bid value ,
	class		= symbols-rank 
}

\DeclareAcronym{rank_bid_node_resources}{
    short   = \ensuremath{c_1} ,
    long    = Node resources criterion for bid assessment ,
    class     = symbols-rank
}

\DeclareAcronym{rank_bid_proximity}{
    short   = \ensuremath{c_4} ,
    long    = Node proximity criterion for bid assessment ,
    class     = symbols-rank
}

\DeclareAcronym{rank_bid_performance_history}{
    short   = \ensuremath{c_5} ,
    long    = Node performance history criterion for bid assessment ,
    class     = symbols-rank
}

\DeclareAcronym{rank_bid_current_node_resources}{
    short   = \ensuremath{c_2} ,
    long    = Node current node resources criterion for bid assessment ,
    class     = symbols-rank
}

\DeclareAcronym{rank_bid_fairness}{
    short   = \ensuremath{c_3} ,
    long    = Fairness criterion for bid assessment ,
    class     = symbols-rank
}

\DeclareAcronym{rank_bid_current_node_resources_unnormalized}{
    short   = \ensuremath{\widetilde{c_4}} ,
    long    = Fairness criterion for bid assessment (non normalized),
    class     = symbols-rank
}

\DeclareAcronym{rank_requirement_cap_i_unnormalized}{
	short	= \ensuremath{\widetilde{\rho_i}},
	long 	= requirement assessment, 
	class	= rank-symbols
}

\DeclareAcronym{datasize}{
    short   = \ensuremath{\lambda_{\text{data}}} ,
    long    = Data size,
    class   = symbols-tas
}

\DeclareAcronym{bandwidth}{
    short   = \ensuremath{W} ,
    long    = Network interface bandwidth,
    class   = symbols-tas
}

\DeclareAcronym{t_transmission}{
    short   = \ensuremath{t_{\text{tx}}} ,
    long    = Time required for transmission of a frame within a data flow of a service,
    class   = symbols-tas
}

\DeclareAcronym{t_needed}{
    short   = \ensuremath{t_{\text{needed}}} ,
    long    = Time required for transmission of a frame within a data flow of a service including guard time,
    class   = symbols-tas
}

\DeclareAcronym{t_guard}{
    short   = \ensuremath{t_{\text{guard}}} ,
    long    = Guard time,
    class   = symbols-tas
}

\DeclareAcronym{t_free}{
    short   = \ensuremath{t_{\text{free}}} ,
    long    = Free unused time within the time the gate for the traffic class is opened,
    class   = symbols-tas
}

\DeclareAcronym{t_free_k}{
    short   = \ensuremath{t^{k_x}_{\text{free}}} ,
    long    =  ,
    class   = symbols-tas
}

\DeclareAcronym{t_free_k_plus}{
    short   = \ensuremath{t^{k_{x+1}}_{\text{free}}} ,
    long    =  ,
    class   = symbols-tas
}

\DeclareAcronym{t_open}{
    short   = \ensuremath{t_{\text{open}}} ,
    long    = Time the gate for the traffic class is opened,
    class   = symbols-tas
}

\DeclareAcronym{traffic_class}{
    short   = \ensuremath{k} ,
    long    = Traffic class within a network interface,
    class   = symbols-tas
}

\newcommand\symb[1]{%
    \acs{#1}%
}

\newcommand\symbrank[1]{%
    \acs{rank_#1}%
}

\newcommand\circled[3]{%
    \tikz[baseline=(char.base)]{
        \node[shape=circle, fill=#1, inner sep=0pt, text width=8pt, align=center]
            (char) {\textcolor{#2}{\normalfont{\sffamily\bfseries\scriptsize #3}}};
    }%
}

\newcommand\bcircle[1]{%													%%% Create white text within a black circle.
    \circled{black}{white}{#1}%
}

\def\checkmark{\tikz\fill[scale=0.4](0,.35) -- (.25,0) -- (1,.7) -- (.25,.15) -- cycle;}

\newcommand{\ie}[0]{\textit{i.e.}~}

\definecolor{royalblue}{rgb}{0.25, 0.41, 0.88}

\begin{document}

\begin{frontmatter}

\title{Self-assessment approach for resource management protocols\\in heterogeneous computational systems\tnoteref{t1}}
\tnotetext[t1]{This work was supported in part by grants from FCT (Grant Number: 2021.06223.BD and 2023.04459.BD).}

\author[1,2]{Rui Eduardo Lopes\corref{cor1}}
\ead{ruieduardo.fa.lopes@ua.pt}
\author[2]{Duarte Raposo}
\ead{dmgraposo@av.it.pt}
\author[1]{Pedro V. Teixeira}
\ead{pedro.teix@ua.pt}
\author[1,2]{Susana Sargento}
\ead{susana@ua.pt}

% Affiliations.
\affiliation[1]{organization={Universidade de Aveiro},
				addressline={Campus Universitário de Santiago},
				city={Aveiro},
				postcode={3810-193},
				country={Portugal}}

\affiliation[2]{organization={Instituto de Telecomunicações},
				addressline={Campus Universitário de Santiago},
				city={Aveiro},
				postcode={3810-193},
				country={Portugal}}

\begin{abstract}
With an ever growing number of heterogeneous applicational services running on equally heterogeneous computational systems, the problem of resource management becomes more essential. Although current solutions consider some network and time requirements, they mostly handle a pre-defined list of resource types by design and, consequently, fail to provide an extensible solution to assess any other set of requirements or to switch strategies on its resource estimation. This work proposes an heuristics-based estimation solution to support any computational system as a self-assessment, including considerations on dynamically weighting the requirements, how to compute each node's capacity towards an admission request, and also offers the possibility to extend the list of resource types considered for assessment, which is an uncommon view in related works. This algorithm can be used by distributed and centralized resource allocation protocols to decide the best node(s) for a service intended for deployment. This approach was validated across its components and the results show that its performance is straightforward in resource estimation while allowing scalability and extensibility. 
\end{abstract}


\end{frontmatter}

\section{Introduction}
Applicational services have several types of execution requirements such as memory or storage, but also network an time requirements, including time sensitive networking (TSN) requirements such as the expected end-to-end or the jitter, that are either statically defined by the service developer or could be estimated. These requirements can be hard or soft requirements, depending on their importance to the overall service operation. 

Within the context of edge or fog computing resource management, features could be critical to allow services to be fully deployed and to display themselves fully available to their own requirements. As the dimension of the network infrastructure, either in number of network nodes, diversity of these types of nodes, and number of services can be quite large, both the placement and the configuration of new applications on them could be troublesome: the efforts required for coordination in resource allocation for services are larger, the higher the dimension is.

We consider, for the sake of simplicity, a scenario where a client, attached to a network node, is requesting content to be delivered to it with ultra-low latency requirements and requiring high computational power, such as in a cloud gaming service. Services such as this one are commonly characterized by having high fluctuations in their content demand (especially in peak hours), and require nodes to allocate CPU and GPU resources dynamically to maintain frame rates and prevent frame drops or resource starvation for lower-priority users. A mechanism able to adaptively allocate resources, capable of configuring time-sensitive measures to tackle latency, and being priority-aware is paramount.

In another perspective, challenged by a more diverse heterogeneity of nodes, in the case of a smart city, multiple different services compete for the same resources (either they are computational, for communication, storage, or any other), strongly varying for their specification of both priority and demand~\cite{Pires2023}.

\begin{figure}[h]
    \centering
    \includegraphics[width=\columnwidth]{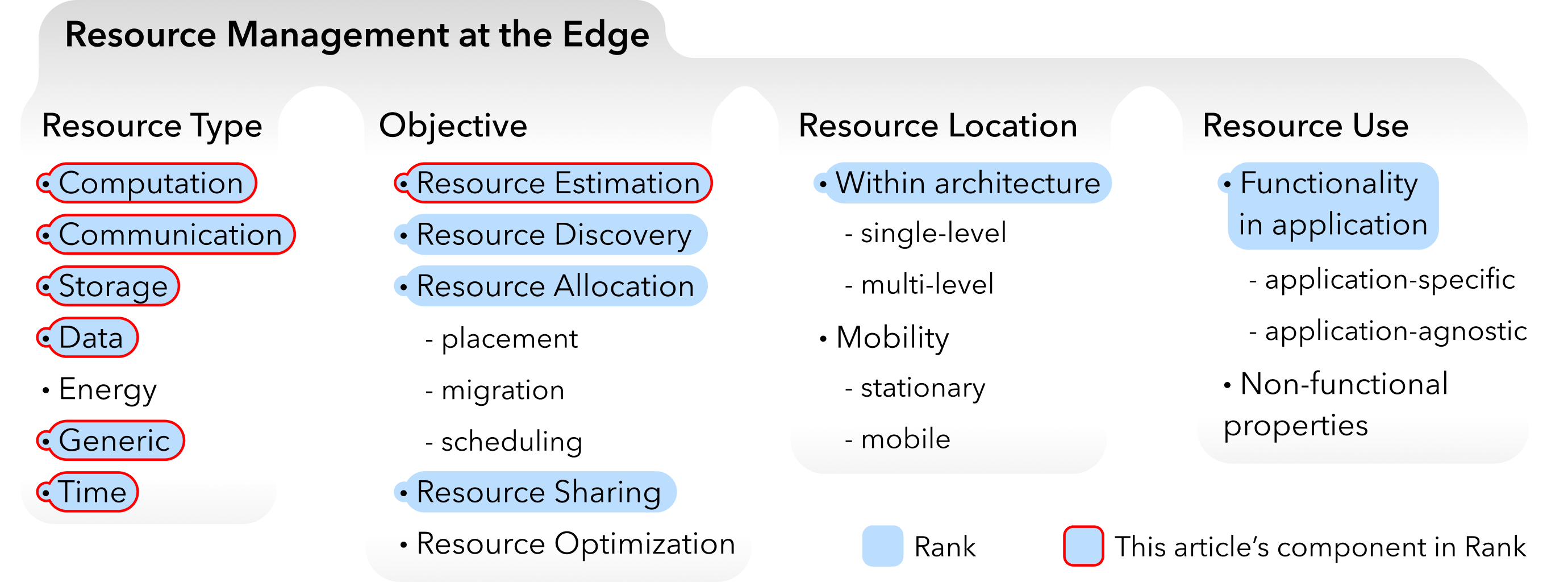}
    \caption[Resource management taxonomy]{Resource management taxonomy (adapted from~\cite{Tocze2018}).}
    \label{fig:res-taxonomy}
\end{figure}

Resource management protocols are responsible for ensuring that the services' requirements are met and that the necessary computational and networking resources are properly managed and allocated. Figure~\ref{fig:res-taxonomy} depicts a taxonomy detailed in~\cite{Tocze2018} for the functionalities and topics that a resource management protocol that works at the network edge level considers. Although it might appear, the exhaustive list of features in figure~\ref{fig:res-taxonomy} does not mean that every example of protocol must consider all items at once. In fact, the combinations for current existent protocols are several. In this article, we focus on our proposed protocol, called Rank, able to consider computation, communication, storage, data, time, and other resource types within the architecture of a given network to provide functionality in applications and services (Rank features are marked in blue in figure~\ref{fig:res-taxonomy}).

From the taxonomy in figure~\ref{fig:res-taxonomy}, Rank is being designed as a fully-distributed resource management protocol able to perform resource estimation, discovery, sharing, and allocation. This protocol works so that admission requests comprised of a list of requirements and a priority level are assessed node-by-node, from a source to a destination node. Looking over other protocols and mechanisms related to resource management tasks, most works lack the support for multiple resource types (only focusing on network or computing features), and have their resource estimation algorithms deepened into their core, incapable of interchange with other strategies for estimation.

As followed by the described related work in section~2, Rank mainly differentiates itself from other protocols by considering a variable set of resource types, which leads up to the question \enquote{how could a node estimate its own available resources for an admission request on such multitude of resource types?}, and by bringing both a new protocol that is able to solve this question, and a resource estimation algorithm that is thought not to be exclusive to Rank's, but also compatible with other protocols and mechanisms that are able to accommodate it.

In this article we propose a simple-yet-effective algorithm for a machine to self-assess its capabilities of providing its own resources to attend a given list of requirements, in a way that a grade can be compared between multiple other machines' assessments with a same input, as an intermediate step of Rank or other fully-distributed resource management mechanism. This approach is able to match the requirements of a request with a suitability grade of a node, considering criteria of node resources, fairness, proximity and performance in a novel and overall algorithm. In figure~\ref{fig:res-taxonomy} the impact of this proposal is outlined in red, which considers the areas that are covered and addressed by Rank in this article.

%% Snipped from INFOCOM's submission %% The proposed approach is validated using an implementation for analytical analysis, showing the evolution of bidding with the metrics, the process of self-organized and distributed tie-break, multiple requirement admission requests, and schedulability of different traffic classes in a time-sensitive network. The results show that the proposed algorithm is a straightforward approach to assess and compute an admission request, with high scalability and open for extension with any future requirements. %\textcolor{red}{More information about the proposal, and its results...}

This paper starts by briefly describing related work in section~2, where by reviewing some resource management network protocols and frameworks, we explore their approaches on resource estimation and highlight our work as novel. From here our proposed solution is described in section~3, covering the issues and its functionalities. Following a careful description of the evaluation criteria of our proposed algorithm, in section~4 we perform the validation of the assessment process we are proposing. Following, in section~5 a variation of the validation procedure was performed so to include the possibility of assessing multiple-requirement requests. To close the validation, in section~6, we specify how new requirement types could be added in this system, as well as an example relative to a request admission evaluation with \acfp{tsn} requirements for a time scheduling. This paper ends with section~7, where some conclusions are taken, along with some lines of future works to be developed from this point.

\section{Related Work}
\label{sec:related-work}

\begin{figure}[h]
    \centering
    \includegraphics[width=\columnwidth]{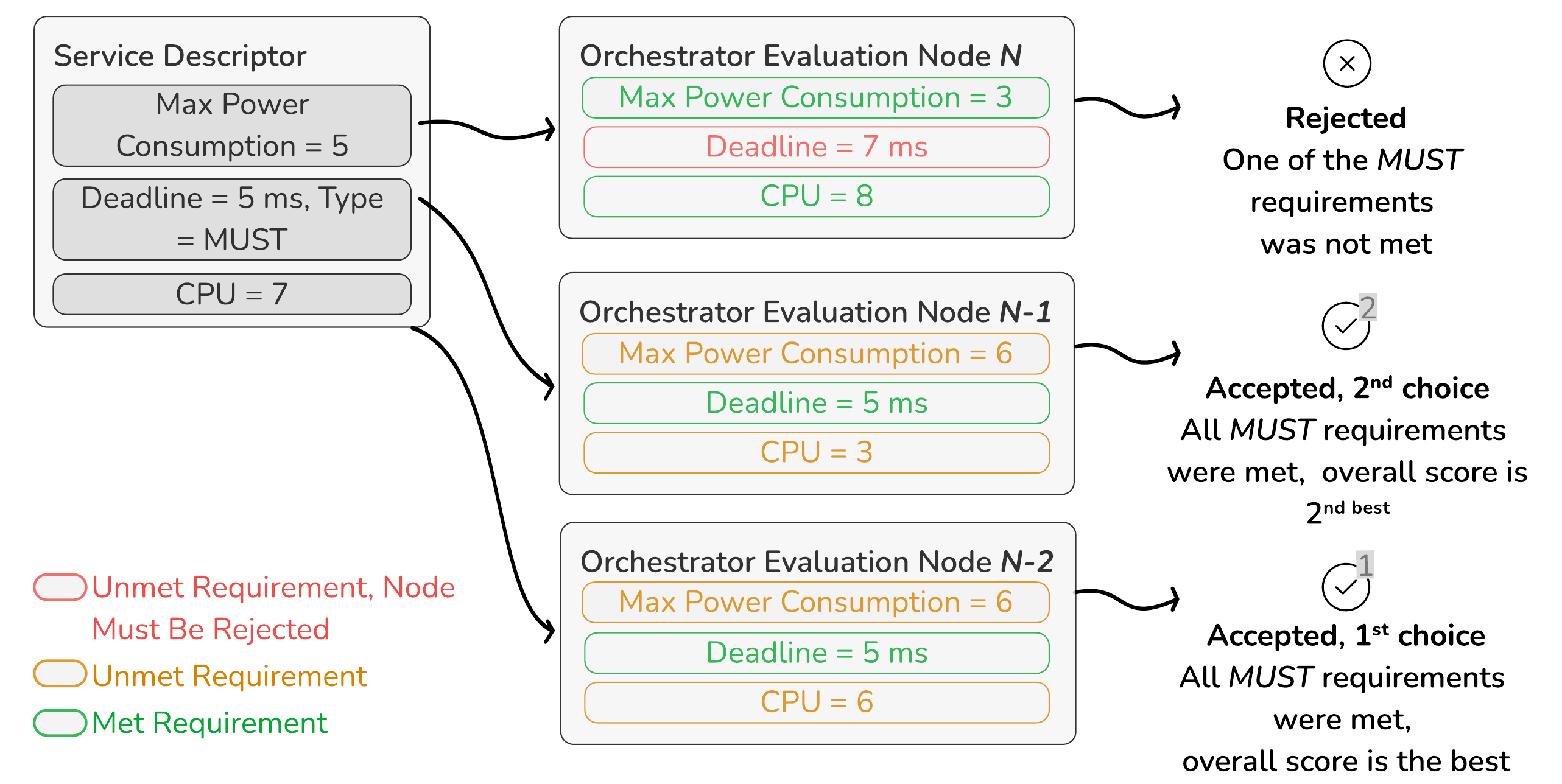}
    \caption{Outcome of evaluation algorithms.}
    \label{fig:related-work-evaluation-goal}
\end{figure}

Within the task of deciding which resources are available, the computational nodes have to be evaluated—either by themselves or by other entities—on their capability to fulfill each requirement. The goal of such evaluation algorithms is to provide an assessment of each node's capabilities to further decide which will have their resources allocated, as depicted in figure~\ref{fig:related-work-evaluation-goal}. Such evaluation shall take into account the application requirements to decide the best nodes to host the application, and the nodes that shall be rejected, and provide that information to the module(s) responsible for the resource discovery and allocation tasks.

The assessment of nodes' capabilities is a wide-known problem in networking~\cite{Peterson2022} applicable under queueing disciplines circumstances, TCP congestion control techniques, cloud computing~\cite{Muhammad2022} or even network slicing~\cite{Sambamoorthy2021}. Despite being a transversal issue, current solutions are based on very specific scenarios of transmission resource management, and a small subset of requirements, usually hard to modify in number~\cite{Yousafzai2017,Arani2020,Rublein2022,Khan2023}.

Different resource management protocols consider different metrics and their own algorithms to evaluate computational nodes resources and their capability to fulfill certain requirements~\cite{Sunilkumar2014,Bachiega2023,Mahapatra2024}. Examples of resource management protocols include the first \acf{rsvp}, the \ac{srp} designated as IEEE~802.1Qat, and, more recently, the \ac{rap} proposal. 

%RSVP
While in \ac{rsvp}, the resource specification has the requirements, a desired \ac{qos}~\cite{rfc2205}, per the structure of integrated service models, %which includes information on the expected traffic's characteristics, such as~\cite{rfc2210}: token bucket size and rate; peak data rate; minimum policed unit; maximum packet size; rate and slack term.
%\cite{238150}. 
%While theoretically good, the soft state means potential scalability issues, specially in networks with many nodes - such as smart cities.
% SRP 
%Within this protocols, the "Multiple Stream Registration Protocol (MSRP) is a signaling protocol that provides end stations with the ability to reserve network resources that will guarantee the transmission and reception of data streams across a network with the requested quality of service. ". Within it
%Another protocol, 
the \ac{srp} (IEEE 802.1Qat) is comprised of a registration protocol and a set of reservation protocols~\cite{IEEE802.1Qat:online,StreamRe57:online}. These protocols consider stream reservation specifications based on: a numerical priority information (similar to VLAN's priority code point (PCP)); a \textit{rank} flag that identifies the emergency of a stream; and other indicators such as of maximum frame size, interval frames, and worst-case latency that is increased by each bridge, propagating its value through the network. Nonetheless, neither protocol allows their estimation algorithms to be exchanged by others, each one assessing the nodes current resources in their own static way.
%\begin{itemize}
%    \item Priority information (to generate a \ac{pcp} field), through a numeric \texttt{data frame priority} value and a boolean value \texttt{rank} to identify emergency vs. non-emergency streams.
%    \item Traffic specification, including \texttt{MaxFrameSize}, \texttt{MaxIntervalFrames} and worst-case latency \texttt{AccumulatedLatency} (which is increased by each bridge it propagates through the network and initialized based on worst case times for internal processing or media access delay).
%\end{itemize}Listeners can provide feedback if they were successful at reserving resources, with the information given in case of failure only providing feedback on insufficient bandwidth, bridge resources, invalid configurations, as a boolean outcome, which invalidates a comparison between computational nodes.

% \begin{figure}
%     \centering
%     \includegraphics[width=\columnwidth]{image2.png}
%     \caption{Failure codes for several scenarios of failure of a reservation of resources. While they provide information on the reason of failure, such information is in a discrete scale and does not provide information on the adequacy of a node \textit{vs} other to provide the wanted requirements.}
%     \label{fig:related-work:srp-codes}
% \end{figure}

%\subsection{IEEE 802.1Qat-2010 Resource Allocation Protocol}

The \ac{rap}, a proposal in IEEE~802.1Q (IEEE~802.1Qdd, still in draft at the time of writing), \enquote{provides support for accurate latency calculation and reporting}~\cite{P8021Qdd97:online}. Within this protocol, a \enquote{cost estimator that can estimate the cost of process executions}, possesses the \enquote{knowledge of local conditions, resources required by a process, and local capacities}. The estimated cost, part of the overall cost, is used to decide the best node~\cite{1602176}. However, it fails to fully consider other resources of the node and their effectiveness in ensuring requirements of a certain service. 

\begin{table*}[ht]
\centering
\caption{Mainly used considered resource types as found in the scientific literature.}
\label{tab:sota-by-resource-type}
\footnotesize
\begin{tabularx}{\linewidth}{lX}
\toprule
Resource Type & References \\ \midrule
Computational & \cite{8815852, 9049050, 8676306, Chen2020, gao2020comddpgmultiagentreinforcementlearningbased, 9197692, 8761190, 9272628, ranadheera2017mobileedgecomputationoffloading, https://doi.org/10.1002/cpe.5162, 8012473, 8519737, ZHOU2020107334, 7307234, 8740949, 8985335, 8422240, 8720039, 8651320, 8802256, 8885368, 8792989, 9059015, 10.1007/978-3-030-37262-0_12, Aazam2016} \\
Communication & \cite{9355609, ranadheera2017mobileedgecomputationoffloading, 7307234, 8740949, 8802256} \\
Energy & \cite{9351561} \\ \bottomrule
\end{tabularx}
\end{table*}

\begin{table*}[ht]
\centering
\caption{Considered resource types by resource estimation features of related protocols and frameworks.}
\label{tab:comparison}
\footnotesize
\begin{tabularx}{\linewidth}{llXXXXXXX}
\toprule
 & Name & \multicolumn{1}{l}{Computation} & \multicolumn{1}{l}{Communication} & \multicolumn{1}{l}{Storage} & \multicolumn{1}{l}{Data} & \multicolumn{1}{l}{Energy} & \multicolumn{1}{l}{Generic} & \multicolumn{1}{l}{Time} \\ \midrule
Protocols & RSVP &  & \checkmark &  &  &  &  &  \\
 & SRP &  & \checkmark &  &  &  &  & \checkmark \\
 & RAP &  & \checkmark &  &  &  &  & \checkmark \\ \midrule
Frameworks & Kubernetes & \checkmark & \checkmark & \checkmark & \checkmark &  & \checkmark &  \\ \midrule
\textit{This Article} & Rank & \checkmark & \checkmark & \checkmark & \checkmark &  & \checkmark & \checkmark \\ \bottomrule
\end{tabularx}
\end{table*}

While it is important to consider network requirements, all previous protocols fail to consider the necessary execution resources (e.g. CPUs, memory, among others)~\cite{Nath2018,Yousefpour2019}. Outside the network protocols domain, container orchestration solutions such as Kubernetes also have to deal with the evaluation of computational nodes against some pre-defined metrics. Kubernetes, within its scheduling, considers auto scaling both horizontally, \ie running multiple instances of a Kubernetes workload, and vertically, \ie resizing CPU and memory resources assigned to containers. With Kubernetes, it is possible to describe the values of a current metric, $m_{\text{current}}$, and a desired metric, $m_{\text{desired}}$, through the definition of target average values~\cite{kubernetes}.%\footnote{Example at When the \texttt{averageValue} or the \texttt{averageUtilization} field in \texttt{target} is specified, $m_{\text{current}}$ is computed by taking the average of the given metric across all Pods, per }.}. 

%\begin{listing}[h]
%\begin{minted}{yaml}
%kind: HorizontalPodAutoscaler
%...
%metrics:
%  - type: Resource
%    resource:
%      name: cpu
%      target:
%        type: Utilization
%        averageUtilization: 50  # %as %$m_{\text{desired}}$%.
%\end{minted}
%\caption{Configuration of horizontal scaling in Kubernetes.}
%\label{listing:related-work:kubernetes}
%\end{listing}

% The horizontal scaling computes the necessary replicas, $D$, as stated in~(\ref{eq:replicas})~\cite{kubernetes}, where $|D|$ means the cardinality of $D$: 

% \begin{equation}
% \label{eq:replicas}
%     |D_{\text{new}}| = \left\lceil |D_{\text{current}}| \times 	\left(\frac{\overline{u}_{\text{current}}}{\overline{u}_{\text{desired}}}\right)\right\rceil
% \end{equation}

The horizontal scaling computes the necessary replicas, as a function of the current replicas times a ratio between the expected average utilization  $m_{\text{desired}}$ of a resource and the real utilization $m_{\text{current}}$. While this solution is more flexible since more types of resources can be considered, and dynamically assign more or less resources depending on a user set/declared requirement, it operates only at the higher service level, without concerns for the underlying network resources. It also does not consider pre-allocation of resources as replicas are created on demand.

In the scientific community there are also other works on resource estimation, most of them covering processes of computational offloading. For this reason, the vast majority of works cover the usage or the focus on this resource type, as shown, in summary, in table~\ref{tab:sota-by-resource-type}.

Besides computational offloading, there are some other usual use cases such as service placement, task allocation, monetary cost usages, or resource usage optimizations~\cite{Ghobaei-Arani2020, Sahni2022, Fahimullah2024}.

% \begin{table*}[ht]
% \centering
% \caption{Main use cases for resource management as found in the scientific literature.}
% \label{tab:sota-by-use}
% \footnotesize
% \begin{tabularx}{\linewidth}{lX}
% \toprule
% Use case & References \\ \midrule
% Offloading & \cite{8815852, 9049050, Chen2020, gao2020comddpgmultiagentreinforcementlearningbased, 9197692, 8761190, 9272628, ranadheera2017mobileedgecomputationoffloading, https://doi.org/10.1002/cpe.5162, 8012473, 8519737, ZHOU2020107334, 7307234, 8740949, 8985335, 8422240, 8720039, 8651320, 8802256, 9059015, 10.1007/978-3-030-37262-0_12} \\
% Scheduling & \cite{9351561, 8676306, 8761190} \\
% Service Placement & \cite{8885368, 8792989} \\ \bottomrule
% \end{tabularx}
% \end{table*}

From what was found, a resource management protocol with allocation purposes should have into its core a more modular resource estimation approach, where nodes are evaluated on the comparison between the expected target and the real current or characteristic value at the node.  Current solutions are missing this flexibility while operating at lower layers. Although there are plenty of solutions for resource allocation protocols in the current state-of-the-art~\cite{Ghobaei-Arani2020, Sahni2022, Fahimullah2024}, not every solution targets the resource utilization as an objective. Our approach, Rank, has this consideration and aggregates resources from a more expanded set of types in relation to others, as described in table~\ref{tab:comparison}.
\section{From Requirements to a Suitability Grade}
\label{sec:architecture}
\label{sec:design}

\begin{figure*}[t]
	\centering
	\includegraphics[width=\linewidth]{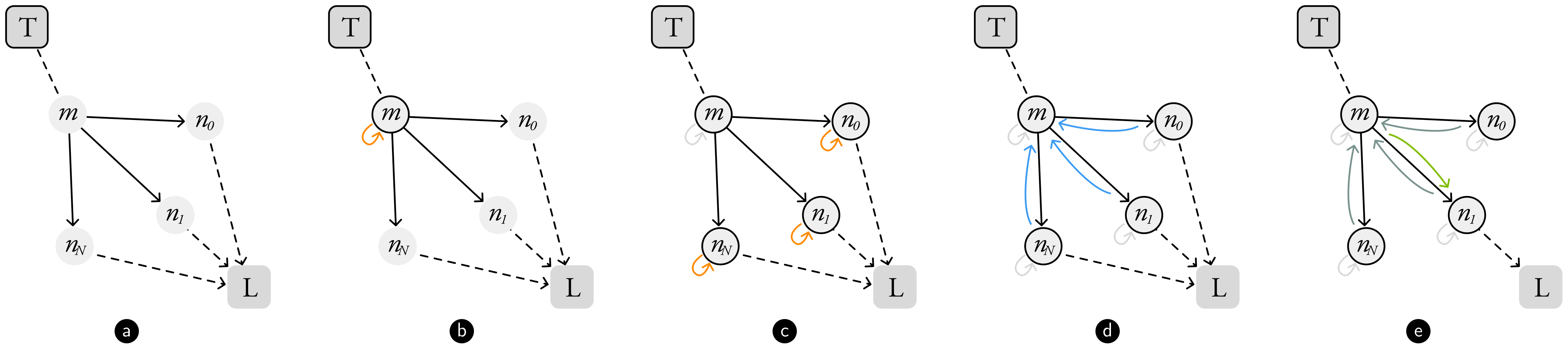}
	\caption{Use of suitability values by a fully-distributed resource allocation protocol, from a talker (T) to a listener (L).}
	\label{fig:bids-chrono}	
\end{figure*}

By mentioning a novel resource allocation protocol that is sensitive to a threefold network, computing, and time set of resources, the ability for a node to assess its capabilities of performing a given service is paramount. To do so, two steps must be taken so that an assessment is well-structured and performed: (i) first, a set of rules must be considered, with a given order and relevance; (ii) then [at time of evaluation of such criteria], the system must be able to successfully retrieve such metrics from itself. 

In a more visual way, Rank, our resource allocation protocol being designed, and hereby considered as a surrogate to this proposed resource self-assessment algorithm, follows a transaction pattern similar to the chronology depicted in figure~\ref{fig:bids-chrono}. Here, a node $m$ received a resource allocation request message forwarded from a talker $T$, and intends to proceed with the request through some node $n_i$ that currently grants a connection to the target listener $L$. In this figure each iteration is being represented by a different stage: in~\bcircle{a} the node $m$ receives the message; in~\bcircle{b} the node $m$ proceeds to self-assess its capabilities of performing the resource allocation request, canceling it if not capable; as node $m$ has three connecting nodes that are able to provide it with a path towards $L$, in~\bcircle{c} node $m$ sent a message to each neighboring node $n_i$, leaving them to self-assess their own capabilities of performing what is being requested; in response to this action, in~\bcircle{d} each node answers back to node $m$ with a suitability value; node $m$, having received the suitability values from every node $n_i$, in~\bcircle{e}, it is now responsible for sorting candidates for forwarding the request message.

In this section, we carefully introduce the steps that are indispensable towards our goal of completing a self-assessment to a service execution capability.

%\subsection{Structure and Design of Criteria for Assessment}
%\label{sec:design}

\begin{table}[t]
\centering
\footnotesize
\caption{List of symbols used in this article}
\label{tab:symbols}
\begin{tabular}{p{0.1\columnwidth}p{0.8\columnwidth}}
\hline
$B$ & suitability value \\ 
$c_x$ & criterium $x$ of assessment \\
$\delta_i$ & historical parameter weight \\ 
$L$ & listener \\
$R$ & list of requirements \\
$\rho_i$ & individual assessment of requirement $i$ \\
$p$ & priority \\
$s$ & salt \\
$\tau$ & requirement weight distribution threshold \\
$T$ & talker \\ 
$\vartheta$ & salt weight \\ \hline
\end{tabular}
\end{table}

\subsection{The structure of an admission request}

Consider a resource allocation protocol that exchanges messages carrying admission requests with a list of requirements (describing a service) and a given priority. When approached by a neighboring node with an admission request carrying both a list of requirements $\symbrank{requirements}$ and a priority $\symbrank{priority}$, a machine should start a procedure for self-assessment towards the goal of performing $\symbrank{requirements}$.

The list of requirements $\symbrank{requirements}$ is comprised by a custom multitude of computational, network, and time resource items, ordered by their intended relevance. This list might be done by an external service descriptor entity or software component that is able to characterize a service into a list of required resources to be successfully performed. This same component might also designate a priority value, $\symbrank{priority}$, that follows $\symbrank{requirements}$ within the admission request, in case it needs to be distinguished among other requests.

For the sake of illustration, for instance, for a given service to be deployed upon an infrastructure, an admission request with requirements of a guaranteed \qty{1}{\giga\byte} of memory, 450~kbps of bandwidth, and capable of allowing a \ac{tsn}-based shaper rule, may be requested to be allocated, with a given priority value. The order of the items in $\symbrank{requirements}$ is relevant, so that the availability to use all the specified requirements is also assessed. In such scenario, if there is at least one requirement that could not be achieved, then the admission request must be automatically denied.

Having a multitude of different resource types as we consider in this proposed assessment means that every listed requirement in $\symbrank{requirements}$ must be graded in a normalized scale. To achieve this, and since not every two requirements are to be assessed in the same fashion, every resource must grant a function to assess themselves in a given feature. In the following subsection we will get into this issue in more detail, where we describe the self-assessment procedure of a node.

\subsection{The procedure for self-assessment}

A receiving node, having to self-assess its capabilities towards the provision of the requested service, must be able to follow a set of criteria to obtain a grade that must be normalized in a way that comparisons regarding these values can be considered inherently valid. For this reason, let us consider a value between $0$ and $1$, in which $0$ means that the node is incapable of providing such a service, whilst $1$ means that the node is fully-capable to perform it.

The procedure of self-assessment within a node against $\symbrank{requirements}$ does not seem to be a difficult execution, but since such a protocol needs to do more than just a current evaluation of the node's resources, and then it needs to ensure that a resource allocation is guaranteed to be working during the interval in which it is in place, several criteria needs to be taken into consideration to declare that a given node is more capable than another in order to perform a given request under a strict list of requirements. The following criteria were designed: 1) \textit{Bare-metal node resources}; 2) \textit{Current node resources}; 3) \textit{Priority}; 4) \textit{Proximity}; and 5) \textit{Historical performance}. At the end, a suitability value, \symbrank{bid} in (\ref{eq:rank-bid-function}), will be computed with these contributions from the following described criteria. 
\begin{equation}
\label{eq:rank-bid-function}
\begin{split}
f\left(R, p\right)=\symbrank{bid} &= \symbrank{bid_node_resources} \times \symbrank{bid_current_node_resources} \times \symbrank{bid_fairness} \times \frac{\symbrank{bid_proximity} + \symbrank{bid_performance_history}}{2}
\end{split}
\end{equation}

%In the next subsections we briefly describe how each criterion allows the assessment to be specialized, ending with how could they contribute to a global grade at the end.

\subsection*{Bare-metal Node Resources, \symbrank{bid_node_resources}}

Considering $\symbrank{requirements}$, the most primitive way one has to check if the resources are capable of performing such specifications is by checking if the surrogate machine has bare-metal components that could support such an admission request. This generates the first criterion of bare-metal node resources, resulting in two possible outcomes: 

\begin{itemize}
	\item $0$ if the node's total resources do not support $\symbrank{requirements}$;
	\item $1$ if the node's total resources do support $\symbrank{requirements}$.
\end{itemize}

If this bare-metal resource assessment fails to successfully grant with a positive outcome, then no more criteria is required to be evaluated, as the current node will never, at any circumstance, be able to successfully perform the admission request.

\subsection*{Current Node Resources, \symbrank{bid_current_node_resources}}

When there are no bare-metal conditions to proceed with an admission request, a refusal is straightforward; otherwise, a tiebreaker must exist to proceed with the evaluation of a node's capability to perform $\symbrank{requirements}$. To perform such an evaluation, it is critical to understand how $\symbrank{requirements}$ is structured. The list $\symbrank{requirements}$ is a collection with relevance the element's insertion order: under the context of $\symbrank{requirements} = \left(a,b,\dots,z\right)$, the first requirement, $a$, is more relevant than the second, $b$, which consequently is highly regarded in relation to the last requirement, $z$. 

The ending self-assessed grade must take into consideration the order with which $\symbrank{requirements}$ is described, which poses the question on how it would be possible to weigh the different requirements according to their position in $\symbrank{requirements}$, while keeping the goal of normalizing such a value to a result in the $[0,1]$ range.

\begin{figure}[h]
	\centering
	\includegraphics[width=\columnwidth]{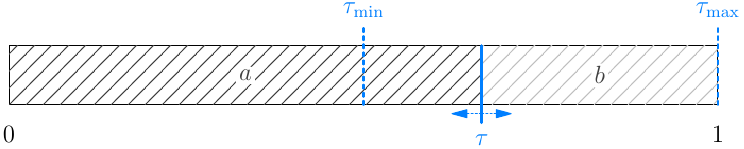}
	%\missingfigure[figwidth=\columnwidth]{the first figure i've drawn with two requirements r1 and r2, and the $\tau$ values}
	\caption{Generic weights distribution between two requirements.}
	\label{fig:two-requirements}	
\end{figure}

If $\symbrank{requirements}$ has a single requirement, then the self-assessment is trivial, since it is its direct evaluation, represented as $\symbrank{requirement_cap}_0$. Following this case, the requirements should weigh different values between them, in a descending order from the first to the \mbox{$i$-th} element of $\symbrank{requirements}$. In figure~\ref{fig:two-requirements} it is depicted a generic weight distribution between two requirements. A given weight $\symbrank{bid_div_constant}$ is applied to the first requirement. Following this reasoning, the second requirement receives the remainder weight of $1-\symbrank{bid_div_constant}$ (since the total assessment value must be 1, that is, in the best-case when all the assessed requirements receive the highest grades $\hat{\symbrank{requirement_cap}_i}$, the total must result in 1: $\hat{\symbrank{requirement_cap}} = \hat{\symbrank{requirement_cap}_0} + \hat{\symbrank{requirement_cap}_1} = 1$). In order to allow the distinction between both stated requirements in $\symbrank{requirements}$, and respecting that the $i$-th requirement has always more value than its following $(i+1)$-th requirement, the value of $\symbrank{bid_div_constant}$ cannot be lower or equal to 0.5 (half its individually-graded assessment value) in $\symbrank{bid_div_constant_min}$, and it cannot be higher or equal than 1 (the individually-graded assessment value) in $\symbrank{bid_div_constant_max}$.

\begin{figure}[h]
	\centering
	\includegraphics[width=\columnwidth]{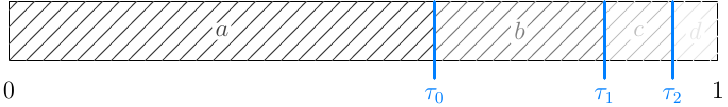}
	\caption{Generic representation of weights distribution between four requirements in a list of requirements $\symbrank{requirements}$.}
	\label{fig:rank-current-node-resource-four-assessment}	
\end{figure}

The choice of the $\symbrank{bid_div_constant}$ value should be parameterizable per node, and is a percentage applied over the individually-graded assessment value of a requirement, $\symbrank{requirement_cap}$. For instance, if the list of requirements has four items $\symbrank{requirements} = \left(a,b,c,d\right)$, for the weights distribution, the first weight is $\symbrank{bid_div_constant}$ towards $\symbrank{requirement_cap}_0$ (the individual assessment of the first requirement), the second weight is $\symbrank{bid_div_constant}'$ towards the $\symbrank{requirement_cap}_1$, and the third weight is $\symbrank{bid_div_constant}''$ towards the $\symbrank{requirement_cap}_2$ (depicted in figure~\ref{fig:rank-current-node-resource-four-assessment}).

Profiting from the recursive behavior depicted in figures~\ref{fig:two-requirements} and~\ref{fig:rank-current-node-resource-four-assessment}, one can write the $\symbrank{bid_function}$ function as in~(\ref{eq:rank-function-n-requirements}).
\begin{equation}
\label{eq:rank-function-n-requirements}
\symbrank{bid_current_node_resources} = \symbrank{bid_function} = \left\{ \begin{array}{ll}      
      \symbrank{requirement_cap}_0 & ,\text{if } \#\symbrank{requirements} =1\\
      \symbrank{bid_div_constant} \cdot \symbrank{requirement_cap}_0 + \left(1-\symbrank{bid_div_constant}\right)\cdot\symbrank{bid_function_reduction} & ,\text{if } \#\symbrank{requirements}>1 \\
\end{array} \right.
\end{equation}

The possibility of parameterizing the value of $\symbrank{bid_div_constant}$ in the range of $(\symbrank{bid_div_constant_min}, \symbrank{bid_div_constant_max})$ allows a node to control a subjective amount of considered requirements. If this weight value is more proximate to its minimum bound, $\symbrank{bid_div_constant} \rightarrow \symbrank{bid_div_constant_min}$, more requirements within the list of requirements can be considered; otherwise, when our weight value is closer to its maximum bound as $\symbrank{bid_div_constant} \rightarrow \symbrank{bid_div_constant_max}$, only one requirement is widely considered.

By keeping the criterion of insertion order in the list of requirements, the assessing nodes are informed of which requirements should be checked before others. Nonetheless, there is a critical condition that must be enforced: if any requirement assessment $\symbrank{requirement_cap}_{i}$ is equal to $0$, then the entire criterion of the current node resource must yield the same outcome. This must be enforced since a node could have resources to support a given requirement request, but not at the time of the request, meaning that this criterion must be able to cancel out such an admission request and further requirements assessment. In sum, the outcome of this criterion is as follows:

\begin{itemize}
	\item $0$ if any of the individually-assessed requirement $\symbrank{requirement_cap}_{i}$ is $0$; 
	\item otherwise, is the result of $\symbrank{bid_function}$ as defined in~(\ref{eq:rank-function-n-requirements}).
\end{itemize}

\subsection*{Priority, \symbrank{bid_fairness}}

With both bare-metal and current node resources criteria, one can already have a somehow clear vision over how a node is able to assess itself over a given service request. However, the admission request not only carries a list of requirements $\symbrank{requirements}$, as it is also accompanied by a priority $\symbrank{priority}$. 

Consider a scenario where a node could potentially sustain an admission request for a list of requirements but, at a given time, there is no availability to support such an execution. The priority of an admission request is a critical concept, bringing a criterion that should be taken into consideration, allowing to order and to re-evaluate capabilities if a priority is higher than the current resource's holders.

The priority level is relative to the admission request, not with each requirement element of the list of requirements $\symbrank{requirements}$. Similarly to the latter criteria, this criterion must be a value from $0$ to $1$, reflecting the level of priority that the given admission request has, where close to $0$ means the lowest priority level, and $1$ means the highest possible —note that in this criterion it is not possible to get $0$ as a result. In sum:

\begin{itemize}
	\item $1$ if the priority of the admission request is maximum;
	\item $x \in (0,1)$ for any other priority value, proportional to $\symbrank{priority}$.
\end{itemize}

\subsection*{Proximity, \symbrank{bid_proximity}}

At this point, the node was already assessed in regard to the resources themselves (both absolute and current resources) and the priority of the admission request. These metrics, nonetheless, might not be enough to solve the possibility of gathering two or more equal assessment values. Moreover, some assumptions are still not considered, such as the position of the assessing node, connection-wise, towards the targeted listener of the admission request. As this estimation is to be solved in a fully-distributed protocol, network considerations must be taken even in the case that no network-type resource requirement is stated in the list of requirements. This can be solved in a new criterion of node proximity.

Considering a criterion for the node proximity towards a targeted listener of a given admission request is a rather subjective concept, since one can drop multiple conclusions to what \enquote{proximity} means. As proximity, one should understand an association of connectivity evaluation functions towards the final destination node of the admission request under test.

To accomplish the goal of this criterion, then a set of network assessment functions must be used, whose final result is to be gathered as the average of all function's outcomes. This requires that all functions operate by retrieving a value in a normalized range from $\left[0,1\right]$. In sum:

\begin{itemize}
	\item $0$ if the network assessment functions averaged 0;
	\item $1$ if the network assessment functions averaged 1;
	\item $x \in (0,1)$ for any other proximity assessment value, as the average of all network assessment functions used.
\end{itemize}

\subsection*{Historical Performance, \symbrank{bid_performance_history}}

Finally, there is a single criterion that should be set to feature some system hysteresis during the execution of the resource allocation protocol handling these admission requests. This being, we introduce historic performance, as a criterion to help on the self-assessment of a node against a list of requirements $\symbrank{requirements}$. Having such an assessment could lead a node to take into consideration recent admission requests, similar to the one's currently requested, and deliver a grade over how good it is such a performance. 

The historical performance for an admission request should be made by leveraging the following variables: 

\begin{enumerate}
\item $\symbrank{bid_history_1}$, ratio of the number of requirements to the average number of requested resource requirements.
\item $\symbrank{bid_history_2}$, percentage of allocated but unused resource reservations.
\item $\symbrank{bid_history_3}$, ratio of how stable are the available resources.
\item $\symbrank{bid_history_4}$, percentage of strict resource reservations over all allowed admission requests.
\end{enumerate}

Adding to this set of variables, a random value should be appended, so that the odds of two nodes presenting the same exact assessment value are severely diminished without compromising the order of magnitude of the total self-assessment final grade and the consequent disposition of this node towards the admission request outcome. To this reason, a fifth contributor to this criterion should be added to generate some salt over the estimated values of $\symbrank{bid_history_1}$, $\symbrank{bid_history_2}$, $\symbrank{bid_history_3}$, and $\symbrank{bid_history_4}$: the salt, $\symbrank{bid_history_salt}$.

The aforementioned variables are designed to be joint by a weighted sum of parameterizable weights $\symbrank{bid_history_weight}_i$, as described in~(\ref{eq:rank-function-history}), where $\sum{\symbrank{bid_history_weight}_i} = 1$.
\begin{equation}
\label{eq:rank-function-history}
\symbrank{bid_performance_history} = \left(1-\symbrank{salt_valorization}\right)\left(\symbrank{bid_history_weight}_1\symbrank{bid_history_1} + \symbrank{bid_history_weight}_2\symbrank{bid_history_2} + \symbrank{bid_history_weight}_3\symbrank{bid_history_3} + \symbrank{bid_history_weight}_4\symbrank{bid_history_4}\right) + \symbrank{salt_valorization}\symbrank{bid_history_salt}
\end{equation}

As also seen in~(\ref{eq:rank-function-history}), the salt is contributing just for entropy reasons and without compromising the main function of the history performance criterion. Thus, it was assigned the $\symbrank{salt_valorization}$, a small percentage weight (below $1\%$), where the remainder is valued over $1-\symbrank{salt_valorization}$ percentage of the total criterion grade.

\subsection*{The final grade, a suitability value, \symbrank{bid}}

\begin{figure}[h]
	\centering
	\includegraphics[width=\columnwidth]{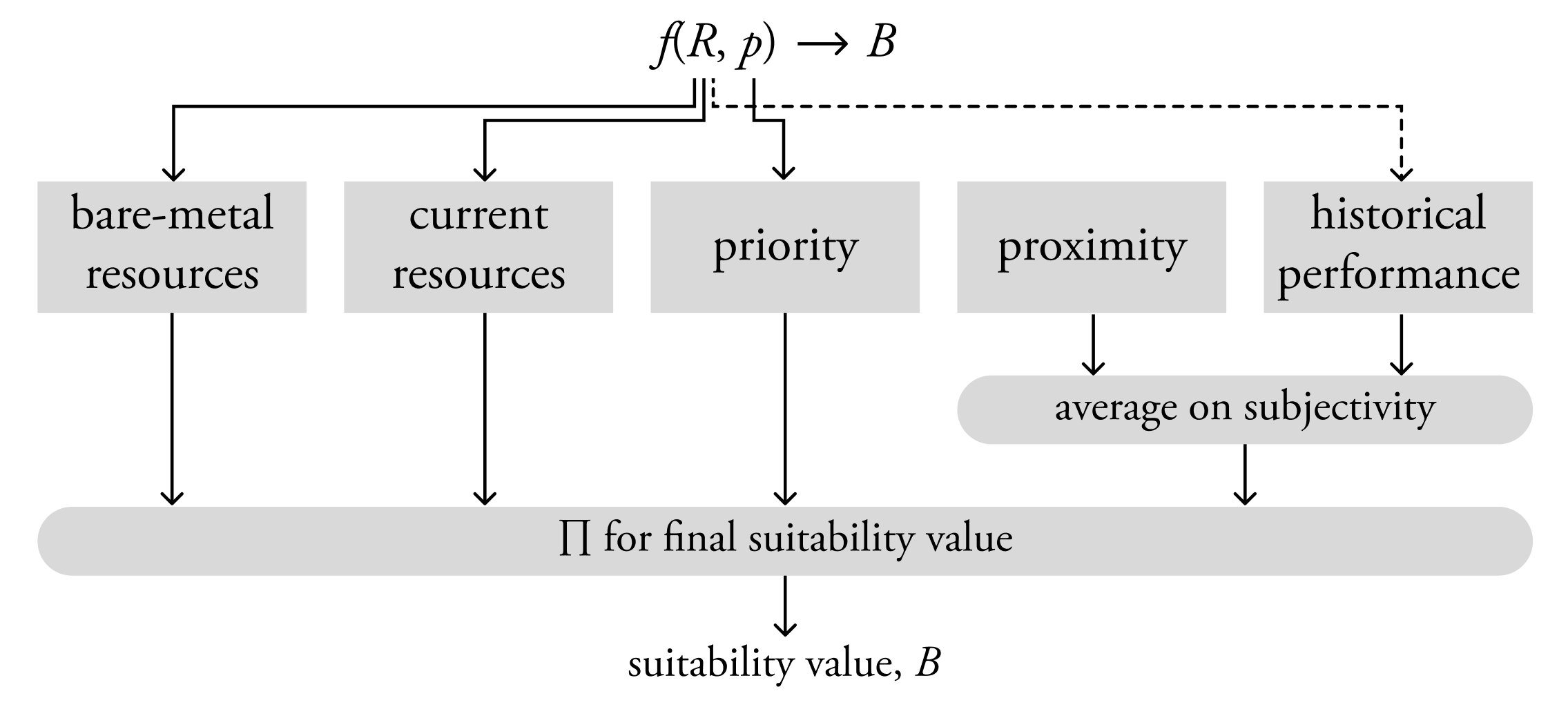}
	\caption{Aggregation of different criteria in a single suitability value. Bare-metal resources and current resources are the only two required resource-related criteria.}
	\label{fig:bid-function}	
\end{figure}

%simply multiplying all the contributors, with the exception of both history performance and proximity criteria, that by being more subjective than the others are to be considered as a contribution via their own average. This resulting value, $\symbrank{bid}$, is called a bid value, since it will be a self-assessing grade that every node receiving an admission request will propose to whom forwarded them such a message, and it is described in
At this point, all the individual contributors to the total and final self-assessment grade are estimated, and a suitability value, $\symbrank{bid}$, can be estimated by~(\ref{eq:rank-bid-function}), and depicted in figure~\ref{fig:bid-function}.

The first three terms in~(\ref{eq:rank-bid-function}), relative to the first three criteria, as they are directly estimated from the list of requirements or the priority given as input, are multiplied between them. The cancellation of one of these three terms must undoubtedly break the resource estimation to zero. The last two criteria, as they are estimated from other parameters rather than what has been given as input, they are considered both as an average, so that they do not cancel out each other. This reflects some subjectivity characteristic of proximity and historical performance criteria.

\section{Evaluation of the Assessment Process}

\begin{figure*}[t!]
	\centering
	\includegraphics[width=\textwidth]{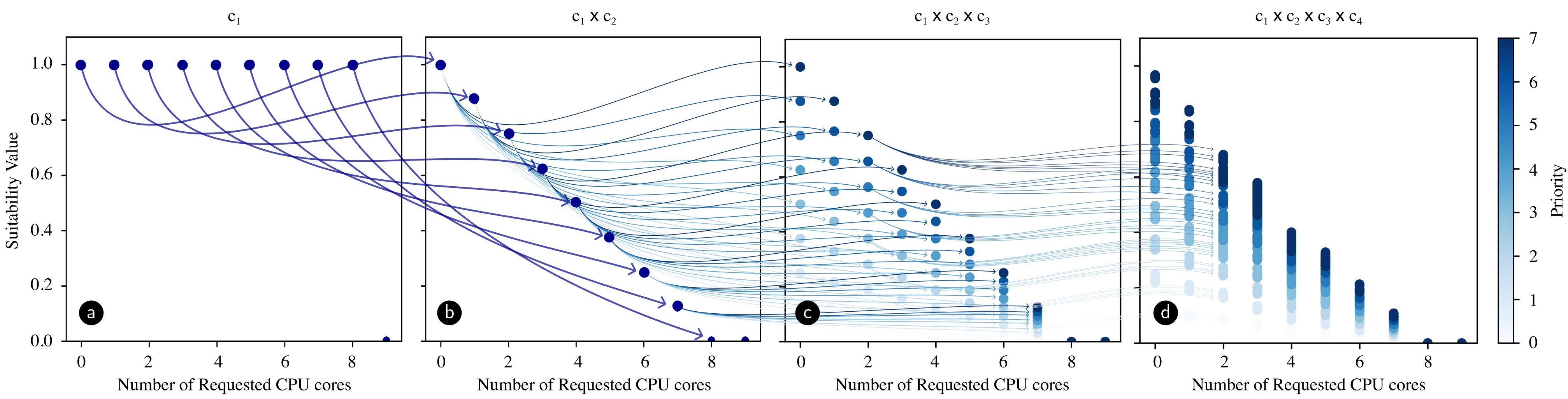}
	\caption{Evolution of assessment by cumulatively add new criteria to its evaluation, diminishing duplicates. \protect\bcircle{a} only bare-metal resources, \symbrank{bid_node_resources}; \protect\bcircle{b} with current node resources, \symbrank{bid_current_node_resources}; \protect\bcircle{c} with priority, \symbrank{bid_fairness}; and \protect\bcircle{d} with proximity criteria, \symbrank{bid_proximity}.}
	\label{fig:bid-evolution}	
\end{figure*}

With the base algorithm for the assessment process already defined, there is the necessity to validate the set of procedures and the impact that each criterion has in the overall assessment grade of a node, towards a list of requirements $\symbrank{requirements}$.

\begin{figure}[h]
	\centering
	\includegraphics[width=\columnwidth]{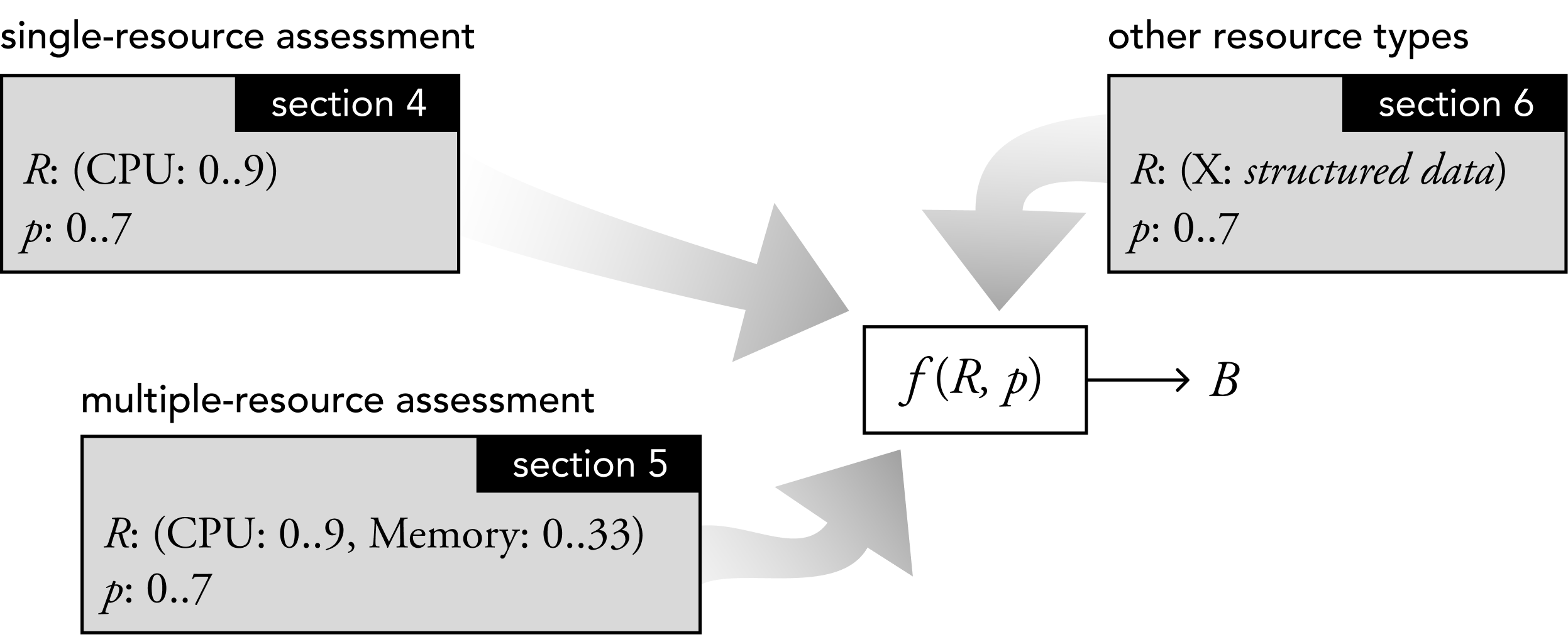}
	\caption{Tests made for validation of our proposed assessment algorithm.}
	\label{fig:bids-tests}	
\end{figure}

\begin{table}[t]
\centering
\footnotesize
\caption{Single-resource assessment evaluation parameters}
\label{tab:experiment-one}
\begin{tabular}{p{0.35\columnwidth}p{0.55\columnwidth}}
\hline
Number of Runs & 100 000 runs \\ \hline
Priority: & 0 to 7 \\
Requirements: &  \\
\quad Requested: &  \\
\qquad CPU & 0 to 9 cores \\
\quad Available: &  \\
\qquad CPU & 8 cores \\ \hline
\end{tabular}
\end{table}

To perform this validation, we assume an easy-to-assess list of requirements $\symbrank{requirements}$ stating a need for a given number of \ac{cpu} cores, as detailed in table~\ref{tab:experiment-one}. The evaluation method will then perform 100~000 requests with a given amount of \ac{cpu} cores, ranging from 0 to 9 required cores, considering that the target machine has only 8 cores in total, 0 being used. This procedure is depicted in figure~\ref{fig:bids-tests} as the section~4 item, followed by the tests described in section 5, and an extension in section~6. All the tests were done by performing the evaluation criteria within a Python script in which machine proximity states were randomized following a uniform distribution.

Figure~\ref{fig:bid-evolution} depicts the suitability assessment in four different phases. From \bcircle{a} to \bcircle{d}, we consider the overall suitability assessment cumulatively considering the bare-metal node resources in \bcircle{a}, the current node resources in \bcircle{b}, the priority in \bcircle{c}, and the proximity in \bcircle{d}. For proximity we considered a four criteria of number of network hops, round-trip time, packet loss, and packet delay variation towards a listener. These conditions were randomly updated throughout the time of the validation execution, following a uniform distribution.

The criterion of node resources only assesses if a requirement would be possible (at anytime) of being accepted in a surrogate node or not. As this criterion does not retrieve any sense of the current capabilities of a node, all the equivalent requests with the same content in $\symbrank{requirements}$ are given the same suitability value, whose result is binary, as seen in figure~\ref{fig:bid-evolution}, phase~\bcircle{a}. Moreover, and as depicted in figure~\ref{fig:bid-evolution}~\bcircle{a}, we can also see that the only case where the suitability value gave out a null value of 0 was when the surrogate machine had, indeed, no capabilities to accept such a list of requirements~$\symbrank{requirements}$. %The criterion of node resources only assesses if a requirement would be possible (at anytime) of being accepted in a surrogate node or not. As this criterion does not retrieve any sense of the current capabilities of a node, all the equivalent requests with the same content in $\symbrank{requirements}$ are given the same suitability value, whose result is binary. 

Adding the second criterion of current node resources, $\symbrank{bid_current_node_resources}$, the suitability values start to take shape according to the current capacities of the surrogate device according to the list of requirements $\symbrank{requirements}$ subject to admission. In figure~\ref{fig:bid-evolution}~\bcircle{b}, one can see the impact of considering these first two criteria in the suitability value.

One can also verify that, in this type of requirement (the \ac{cpu}, as considered for the sake of simplicity and of worst-case analysis), the function behavior of the suitability value in relation to the number of requested cores is clearly linear. Despite this behavior, it is critical to understand that this is not a rule to the estimation algorithm; it is rather a proper behavior defined by the way the evaluation of \ac{cpu} resources (in particular \ac{cpu} core usage) is computed: the evaluation is made through the percentage of the requested value in relation to the number of available cores. The freedom given by this algorithm, to define such a behavior in a per-resource basis, allows us to set safe-guard criteria for over-provisioning scenarios, as one can verify in figure~\ref{fig:bid-evolution}~\bcircle{b}, where requesting 8 cores out of 8 available cores leads to a null suitability value.

Again, in each point of this plot, we still have the 100~000 runs that we have made for each batch of admission requests with a given amount of requested cores. In order to break this tie, we introduce the third criterion of priority, $\symbrank{bid_fairness}$, which is our first level with which we will be able to start distinguishing the batches of admission requests according to given priorities. The effect on appending this criterion is depicted in figure~\ref{fig:bid-evolution}~\bcircle{c}.

The priority level, as assigned by the requester in the admission request, as one can see in figure~\ref{fig:bid-evolution} phase~\bcircle{c}, can hugely impact in the assessment process, directly diminishing the upper bound of the possible values outcome to the evaluation. Through the usage of priorities, it is clear that the maximum suitability value of 1 can only be achieved by admission requests that have the highest priority (defined as being level 7 in a range from 0 to 7). Setting the best individual requirement assessment for the amount of required cores to 0 with priority 6 can only grant the requester with an estimated suitability value of $0.875$, exclusively considering the first three criteria.

Still, and in regard to our effort of attempting to distinguish similar admission requests, we are still unable to clear out duplicate suitability values in this process. From the batches of 100~000 admission requests with different number of requested cores to this latter experiment, we have performed an eighth of the total number of tests in each priority level available. This means that, instead of each point being 100~000 points drawn over others, now we are seeing sets of 125 points carefully overwritten as plotted in figure~\ref{fig:bid-evolution}~\bcircle{c} (an eighth of the total runs for each number of requested cores).

The runs themselves are not distinguishable between them, since there is no criteria being applied to differ a test from another performed in the instant right after. Fortunately, and with the help of the fourth presented criterion of proximity, $\symbrank{bid_proximity}$, this issue can start to be improved. The criteria of proximity states that a component to the evaluation should be estimated through a quite simple evaluation of the current network conditions of the node under test. Considering the base protocol to be working in a distributed manner, it is critical that, even if not requested in $\symbrank{requirements}$, the current network state is reckoned as a downside to the already estimated suitability value. As network conditions are not likely to be repeated in multiple tests, it is believed that the impact of this criteria is sufficient to diminish the chance of duplicate suitability grades. The result of this evaluation in our chronology is depicted in figure~\ref{fig:bid-evolution}~\bcircle{d}.

In regard to the last criterion on performance history, this analysis will be done in two parts: first on how historical parameters create a tendency and are capable of inducing a change in the final estimation; secondly, we will look into the salt value and attempt to reason on both its relevance, and on plausible weight values that one should give to it in the final suitability value, in a way that is sufficient to achieve its objective of diminishing (or deleting) duplicates.

As seen in expression (\ref{eq:rank-function-history}), a first summand is weighted by $\left(1-\symbrank{salt_valorization}\right)$ representing the history performance of admission requests in a given node. These values, as a node starts its operation, will be all started up as zero, since there is no remaining history to allow a node to proceed working on. Alternatively, even when there is already some history, these sub-criteria will not contribute to suitability values being different within, but they are considered to adjust the suitability value in order to add hysteresis to it. This will allow a same node, with the same unchanged set of capabilities, to degrade its suitability value for a same admission request, if there were times in which similar requests did not last long, or were not used as much as the period in which they saved a reservation.

\begin{figure}
	\centering
	\includegraphics[width=\columnwidth]{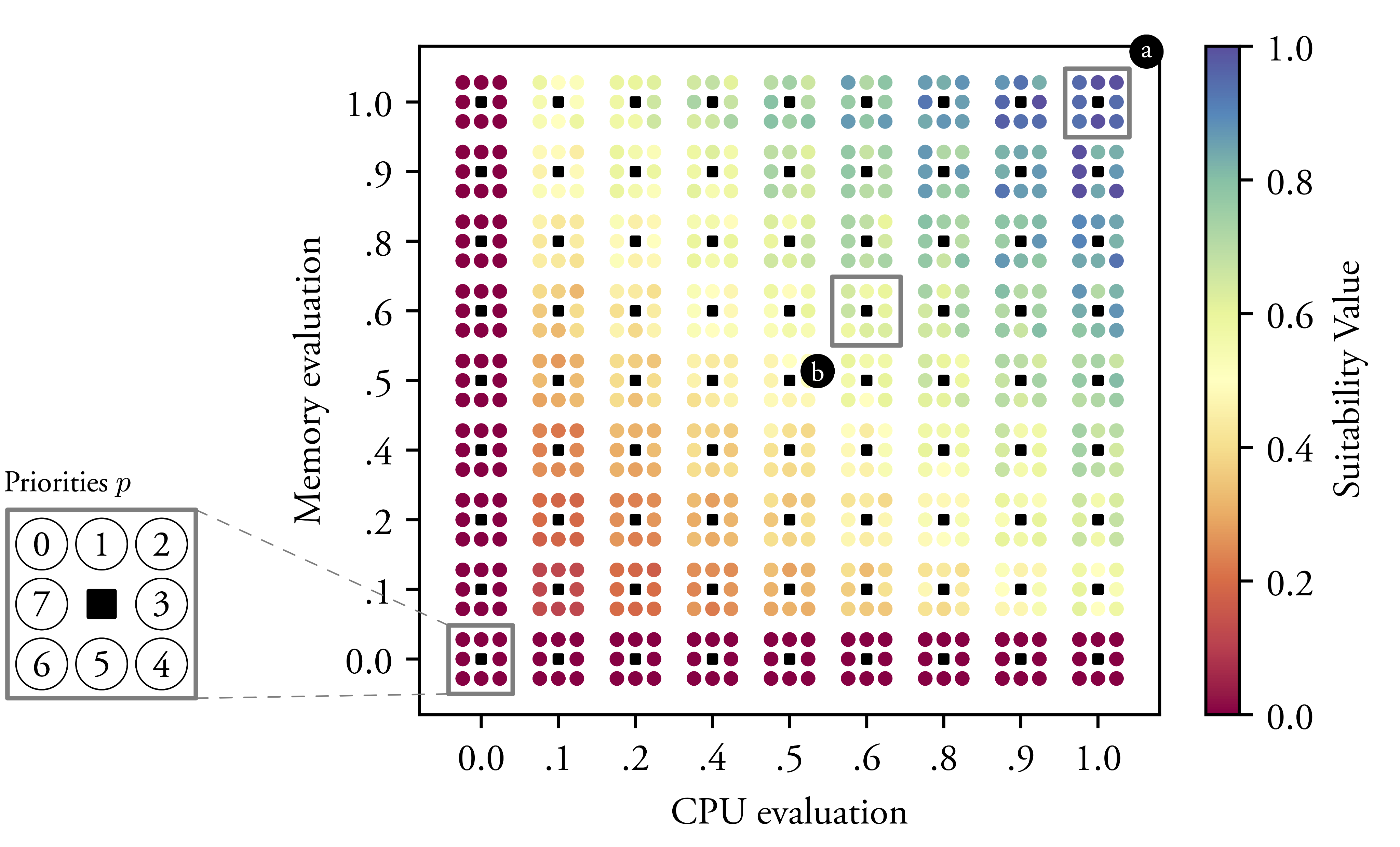}
	\caption{Evolution of suitability values from admission requests of CPU cores and memory, by priority.}
	\label{fig:rank-bid-2req-cpu-mem-0.5}	
\end{figure}

As it is easy to devise from this practice, the variation of values within these sub-criteria of historical performance will not cause suitability values to drastically change throughout time, meaning that they will not be a contribution to diminish duplicate assessments. To this matter, the salt value (as the second summand in (\ref{eq:rank-function-history})) is our last resort to bring some differentiation on every estimated value in an admission request. Given this issue, it is relevant to study the impact of the value of $\symbrank{salt_valorization}$, which is the assigned weight to the salt.

%\begin{figure}[t]
%	\centering
%	\includegraphics[width=.7555\columnwidth]{figs/series_var_salt_main_blue.png}
%	\caption{Evolution of bid value differences with salt weights varying from $[1\times 10^{-20}, 1\times 10^{0}]$.}
%	\label{fig:rank-salt-evolution}	
%\end{figure}

In more algebraic sense, let us consider two equally sourced suitability values as an original value and a salt-added value, $\symbrank{bid}_o$ and $\symbrank{bid}_s$, respectively. Our goal is to adjust the weight $\symbrank{salt_valorization}$, so that randomness is generated between suitability values, but slightly enough not to cause any \textit{drastic} changes, as $\symbrank{bid}_o \neq \symbrank{bid}_s  : \symbrank{bid}_o - \symbrank{bid}_s \approx 0$.

Following this last expression, it is now relevant to better define what we aim by avoiding \textit{drastic} changes. It is important that the added randomness in this system does not affect the evaluation made by the remainder of its estimation as defined in~(\ref{eq:rank-bid-function}). In order to better ensure this, a study was conducted in which we have made the suitability weight to be diminished at a point where its precision simply could not bother a difference to the suitability value. From a salt weight of $\symbrank{salt_valorization} = 1$ (meaning $100\%$), we have dropped its value to $10^{-20}$, and accompanied the results on how the difference $\symbrank{bid}_o \neq \symbrank{bid}_s$ was behaving towards our goal. In table~\ref{tab:rank-salt-summary} one can see depicted this study, in which it is possible to conclude that a weight of $\symbrank{salt_valorization}=1\times 10^{-10}$ is the minimum value possible assignable in which the complete range of valid suitability outcomes is valid: in lower values to $1\times 10^{-10}$, the amount exceeds the precision from the floating point representation.

With this analysis, it is clear that the chosen value for $\symbrank{salt_valorization}$ is dependent of a given real numerical representation. Results in table~\ref{tab:rank-salt-summary} consider  IEEE~754 simple precision floating point representation (chosen for high precision with reasonable occupation within a message). This justifies the difference between bland- and salty-suitability-values dropping to $0$ or even \textit{not-a-number} in the ranges of salt weight $\symbrank{salt_valorization}$ in $[10^{-20}, 10^{-10})$, as summarized in table~\ref{tab:rank-salt-summary} for the validation results. It is zero when it covers subnormal values for the difference of suitability values, and \textit{not-a-number} when it exceeds the representation's precision.

As shown, without placing a non-nullable weight to a salt, the percentage of duplicates was of $14\%$. In a more practical sense, applying this scenario means that a node is able to forward an admission request via multiple other nodes towards an aimed listener, in which duplicates have the probability of occurring by $14\%$, for each pair node-next-node.

\begin{table}
\caption{Summary of salt weights ($\symbrank{salt_valorization}$) evolution and impact on the difference of bids (values are approximations).}
\label{tab:rank-salt-summary}
\scriptsize
\begin{tabularx}{\columnwidth}{lXXXX}
\toprule
\symbrank{salt_valorization} & Minimum & Maximum & Mean & Std. Dev.  \\ \midrule
$1\times 10^{0}$     & $1.80 \times 10^{-7}$  & $4.99\times 10^{-1}$ & $7.91\times 10^{-2}$ & $8.35\times 10^{-2}$ \\
%$1\times 10^{-1}$     & $1.79 \times 10^{-8}$ & $4.99\times 10^{-2}$ & $7.91\times 10^{-3}$ & $8.35\times 10^{-3}$  \\
$1\times 10^{-2}$     & $1.81 \times 10^{-9}$ & $4.99\times 10^{-3}$ & $7.91\times 10^{-4}$ & $8.35\times 10^{-4}$  \\

$\cdots$     & $ $  & $ $ & $ $ & $ $ \\

%$1\times 10^{-8}$     & $1.81 \times 10^{-15}$ & $4.99\times 10^{-9}$ & $7.91\times 10^{-10}$ & $8.35\times 10^{-10}$  \\
$1\times 10^{-9}$     & $1.80 \times 10^{-16}$ & $4.99\times 10^{-10}$ & $7.91\times 10^{-11}$ & $8.35\times 10^{-11}$  \\
\rowcolor{blue!7!} \textcolor{royalblue}{$1\times 10^{-10}$}     & $2.08 \times 10^{-17}$ & $4.99\times 10^{-11}$ & $7.91\times 10^{-12}$ & $8.35\times 10^{-12}$  \\
\textcolor{black}{$1\times 10^{-11}$}     & \textcolor{gray}{$0 \times 10^{0}$}     & $4.99\times 10^{-12}$ & $7.91\times 10^{-13}$ & $8.35\times 10^{-13}$  \\

\textcolor{black}{$\cdots$}     & $ $  & $ $ & $ $ & $ $ \\

\textcolor{black}{$1\times 10^{-18}$}     & \textcolor{gray}{$0 \times 10^{0}$}  & \textcolor{gray}{$0\times 10^{0}$}          & \textcolor{gray}{$0\times 10^{0}$}      & \textcolor{gray}{$0\times 10^{0}$} \\
%\textcolor{black}{$1\times 10^{-19}$}     & \textcolor{gray}{$0 \times 10^{0}$}  & \textcolor{gray}{$0\times 10^{0}$}          & \textcolor{gray}{$0\times 10^{0}$}      & \textcolor{gray}{$0\times 10^{0}$} \\
\textcolor{black}{$1\times 10^{-20}$}     & \textcolor{gray}{\textit{NaN}}  & \textcolor{gray}{\textit{NaN}} & \textcolor{gray}{\textit{NaN}} & \textcolor{gray}{\textit{NaN}} \\ \bottomrule
\end{tabularx}
\end{table}

\section{Assessment on Multiple-requirement Request}

%\pedro{according to Duarte, the first 3 paragraphs could be shortened }
With the latter analysis, a parameter variation is still lacking validation: the multiplicity of requirements within a list of requirements, $\symbrank{requirements}$, as well as its order within such a structure, and the effect it has with the admission request priority. We consider requirements of \ac{cpu} cores and memory, and $\symbrank{priority}$ as the priority of the admission request.

\begin{table}[t]
\centering
\footnotesize
\caption{Multiple-resource assessment evaluation parameters}
\label{tab:experiment-two}
\begin{tabular}{p{0.35\columnwidth}p{0.55\columnwidth}}
\hline
Number of Runs & 1 000 runs per request set\\ \hline
Priority: & 0 to 7 per request set \\
Requirements: &  \\
\quad Requested: &  \\
\qquad \textcolor{gray}{\textit{first request set}} & \\
\qquad CPU & 0 to 9 cores \\
\qquad Memory & 0 to \qty{33}{\giga\byte} \\
\qquad \textcolor{gray}{\textit{second request set}} & \\
\qquad Memory & 0 to \qty{33}{\giga\byte} \\
\qquad CPU & 0 to 9 cores \\
\quad Available: &  \\
\qquad CPU & 8 cores \\
\qquad Memory & \qty{32}{\giga\byte} \\ \hline
\end{tabular}
\end{table} 

The validation will run batches of simulated admission requests (table~\ref{tab:experiment-two}) completing the entire set of combinations for a node with the following procedure: perform 1~000 requests with a) a given amount of required cores (ranging from 0-9), and a given amount of memory (ranging from 0-\qty{33}{\giga\byte}); b) a priority (ranging from 0-7). The requests target simulated machine has 8 cores and \qty{32}{\giga\byte} of memory, in total, none of them being used. Following a combination set of experiments with the \mbox{\acs{cpu}-memory} order, the inverse order (\mbox{memory-\acs{cpu}}) is also tested using the same testing conditions.

Running the suitability estimation for a first requirement list $\symbrank{requirements}_A : \left(\symbrank{requirements}_0, \symbrank{requirements}_1\right)$ and a variable priority $\symbrank{priority}$, means that somewhere in the suitability evaluation algorithm, both order and weight applied to the evaluations of $\symbrank{requirements}_0$ and $\symbrank{requirements}_1$ will be considered, inducing a matching transforming pair $\left(\symbrank{requirements}_0, \symbrank{requirements}_1\right) \rightarrow \left(\symbrank{requirement_cap}_0, \symbrank{requirement_cap}_1\right)$, in which $\symbrank{requirement_cap}_0$ is the capability assessment grade of the $\symbrank{requirements}_0$ requirement, and $\symbrank{requirement_cap}_1$ is the capability assessment grade of the $\symbrank{requirements}_1$ requirement.

For each value of $\symbrank{priority}$ from 0 to 7, figure~\ref{fig:rank-bid-2req-cpu-mem-0.5} depicts a fully-dimensional plot in which one can see how \ac{cpu} and memory assessments impact in the suitability value according to their grades and their request's assigned priority. Notice that each value of requirement assessment is also normalized. In figure~\ref{fig:rank-bid-2req-cpu-mem-0.5}, results for the suitability assessment are counted in bins strategically displaced around centroids locating the suitability value regarding their normalized first requirement assessment (\ac{cpu} number of cores assessment), and their normalized second requirement assessment (memory usage assessment). Analyzing this figure, one can rapidly conclude that it is only needed one requirement assessment to give out a null grade to completely turn the whole suitability analysis nullable. 

In closer inspection, there are some details that might seem counter-intuitive at a first sight. Contrarily to what one would have expected, with this algorithm, having the best grade possible concerning each requirement assessment does not grant the best outcome to a suitability value, as well as a priority. This is due to the network assessment done, whether network capabilities are taken as requirements in the list of requirements or not: since this is supposed to be used in a network protocol, this makes the relevance of the proximity criterion in the suitability evaluation, and allows a protocol to always consider the current network environment, no matter network requirements are declared in an admission request or not.

If we look onto~\bcircle{a} in figure~\ref{fig:rank-bid-2req-cpu-mem-0.5}, it is possible to check that, even in the best of cases (concerning the current capabilities of the node and the admission request priority criteria), if the proximity criterion assesses bad network conditions, the suitability outcome will be affected; this is the reason why we can have suitability values assigned to different priority levels and capabilities not in such a linear and expected way. A more visible case can be seen closer to the center of the plot, in~\bcircle{b}, where the color changes (different suitability values within the color scale accompanying the graph) are clearer.

Another detail that we can analyze is the way the color is changing from bad requirement assessments to the best achieved values. In more detail, it seems that the way colors change and evolve in the $xx$ axis is similar to the same tendency over $yy$, which would be of great unfortunate, since it would mean that our resolution of keeping order in the list of requirements $\symbrank{requirements}$ is not producing the expected results. But, in a closer inspection, and since we still have not taken any consideration onto it until now, this actually might be the output of an evaluation in which the $\symbrank{bid_div_constant}$ threshold is closer to $\symbrank{bid_div_constant_min}$, which is the case.

\begin{figure}[h]
	\centering
	\begin{minipage}[t]{.32\columnwidth}
  		\centering
  		\includegraphics[width=\columnwidth]{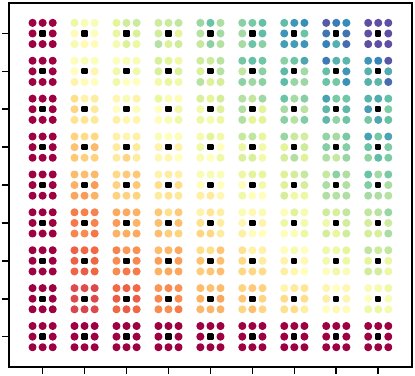}
  		\subcaption{$\symbrank{bid_div_constant} = 0.51$}
	\end{minipage}%
	\hspace{0.005\columnwidth}
	\begin{minipage}[t]{.32\columnwidth}
  		\centering
  		\includegraphics[width=\columnwidth]{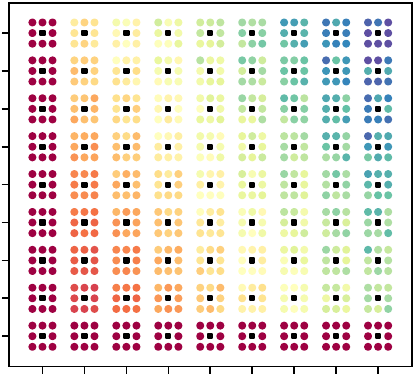}
  		\subcaption{$\symbrank{bid_div_constant} = 0.66$}
	\end{minipage}%
	\hspace{0.005\columnwidth}
	\begin{minipage}[t]{.32\columnwidth}
  		\centering
  		\includegraphics[width=\columnwidth]{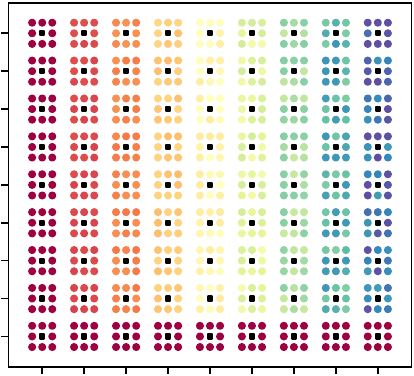}
  		\subcaption{$\symbrank{bid_div_constant} = 0.99$}
	\end{minipage}	
	\vspace{0.25em}
	\caption{Impact of $\symbrank{bid_div_constant}$ threshold in suitability values of multiple-requirement admission requests (axes are similar to figure~\ref{fig:rank-bid-2req-cpu-mem-0.5}).}
	\label{fig:rank-bid-2req-cpu-mem-all}	
\end{figure}

This study on the effect of moving the $\symbrank{bid_div_constant}$ threshold closer to $\symbrank{bid_div_constant_min}$ or to $\symbrank{bid_div_constant_max}$ has visible impact in these validating scenarios, as one can observe in figure~\ref{fig:rank-bid-2req-cpu-mem-all}. In this figure, one can see three different runs differing only in the value of $\symbrank{bid_div_constant}$, which was set to $0.51$ in (a), $0.66$ in (b), and $0.99$ in (c).

From figure~\ref{fig:rank-bid-2req-cpu-mem-all}, it is possible to check the impact that $\symbrank{bid_div_constant}$ has in allowing more or less requirements to perdure their effect in the global suitability outcome. When $\symbrank{bid_div_constant}$ is closer to its minimum possible value, $\symbrank{bid_div_constant_min}$, as it is depicted in figure~\ref{fig:rank-bid-2req-cpu-mem-all}~(a), the most requirements possible are seen as plausible contributors to the final value; in absolute contrast, when $\symbrank{bid_div_constant}$ is closer to its maximum value possible, $\symbrank{bid_div_constant_max}$, no more than a single requirement has impact in the global suitability outcome. This number of assessed requirements by a set of $\symbrank{bid_div_constant}$ values is able to be recognized by relativizing the color variation along the $xx$ axes (which refers to the first requirement declared in $\symbrank{requirements}$), with the color variation along the $yy$ axes (referring to the second requirement declared in $\symbrank{requirements}$). 

Although these examples come from a list of only two requirements set in the admission request, generalizing from this study, we can conclude that $\symbrank{bid_div_constant}$ can be used as a measurement variable definition of exigence of a node. This considers how rigorous a node can be concerning a list of requirements $\symbrank{requirements}$, received as an admission request.

\section{Considering New Resource Types}
\label{sec:assessment}

What if one wants to implement a new and more complex resource type to be assessed by this algorithm? Following the discussion in section~\ref{sec:related-work}, one of the key advantages of this algorithm in relation to other related works is the flexibility we have not only to consider a large array of resource types, but also to be able to expand such an horizon to newer types by demand. Implementing the previously defined self-assessment process into a distributed resource allocation protocol will demand that each considerable requirement has a proper assessment function, since the capability
evaluation of some requirements is completely different (e.g. number of CPU cores \textit{vs} memory usage). Two functions should be designed to give an implementation on the first criterion of bare-metal resources, $\symbrank{bid_node_resources}()$, and other to implement how should a specific requirement be assessed, $\symbrank{requirement_cap}_i()$, as a contributor to \symbrank{bid_current_node_resources}.
%In the latter section, we have explored how is a list of requirements within a admission request assessed (followed by a priority level) and, for the sake of simplicity, we have opted to use easy-to-understand requirement assessments being the number of requested \ac{cpu} cores and memory usage. Implementing this self-assessment and bidding process into a distributed resource allocation protocol will demand that each considerable requirement has a proper assessment function, since the capability evaluation of a given requirement (let us consider the number of \ac{cpu} cores, for instance) must be seen as a completely different task than the evaluation of another requirement (as memory usage). 

%With this described, it is relevant to identify that by each compliant requirement within the distributed resource allocation protocol that would use this self-assessment procedure, a matching assessment interface needs to be implemented. In fact, and regarding the criteria as described before, 

When considering the requirement number of cores, $\symbrank{bid_node_resources}()$ is defined as a function that returns 0 when it is higher than the number of total cores, or is a negative number of cores; otherwise 1 is returned. On the other hand, $\symbrank{requirement_cap}_i()$ is defined as a function that returns a ratio of the requested value to the number of currently available cores. For the memory usage requirement, $\symbrank{bid_node_resources}()$ returns a 0 when it is not physically possible to cling to the requested quantity of memory; otherwise it returns 1, while $\symbrank{requirement_cap}_i()$ returns a ratio of the requested level to the number of currently available memory level.

Although different, both cores and memory usage requirements are quite similar in terms of their evaluations, but this is far from being a more global scenario. To contrast this, let us consider a more complex case of a network \ac{tas} requirement. \Ac{tas} is a time-strict shaper from a set of protocols and mechanisms that comprises what is called a \acf{tsn}. In this shaper one associates (i) the data to traffic classes, and (ii) these to gates that open and close at specific times, and (iii) for specific intervals of time. This ensures no unwanted competition between traffic classes that should not compete for transmission, therefore eliminating the jitter associated with that competition. Within this requirement assessment, the impact of configuring points (i) to (iii) need to be taken into account.

\begin{figure}
    \centering
    \includegraphics[width=\columnwidth]{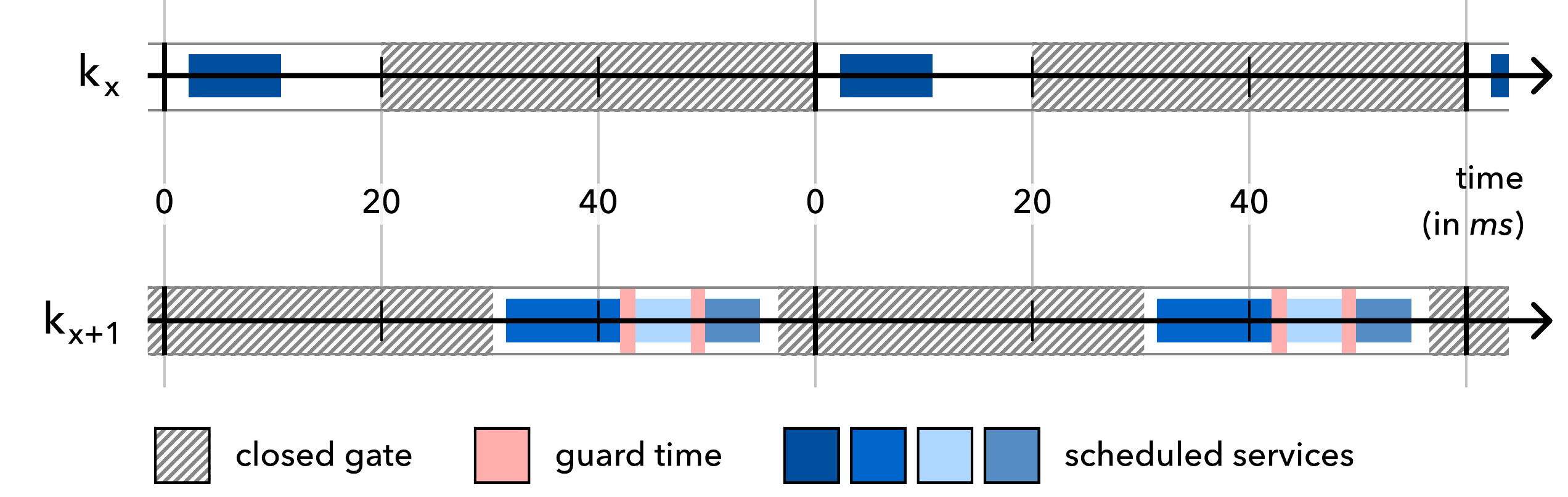}
    %\missingfigure{initial schedule. include values of T open, Tfree and service transmission times}
    \caption{Initial schedule with two traffic classes: $\symb{traffic_class}_{x}$ and $\symb{traffic_class}_{x+1}$.}
    \label{fig:tas-before}	
\end{figure}

%pedro
Considering a new service to be deployed at a set of machines, which is described in terms of its message size \symb{datasize}, and assuming that each node is capable of providing information of its network interface bandwidth \symb{bandwidth}, one can define the transmission time for a data flow of a service, \symb{t_transmission} as (\ref{eq:transmission-time}):
\begin{equation}
    \label{eq:transmission-time}
    \begin{aligned}
        \symb{t_transmission} = \frac{\symb{datasize}}{\symb{bandwidth}}
    \end{aligned}
    \raisetag{20pt}
\end{equation}

When adding a new service, the time needed for transmission \symb{t_needed} is $\symb{t_transmission} + \symb{t_guard}$, where \symb{t_guard} is a guard time to minimize cascading delay issues~\cite{Kim2021}.
%\begin{equation}
%    \label{eq:tneeded}
%    \begin{aligned}
%    \symb{t_needed} = \symb{t_transmission} + \symb{t_guard}
%    \end{aligned}
%    \raisetag{20pt}
%\end{equation}

Within each network interface, a \ac{tas} schedule can be defined for $X$ traffic classes. For each traffic class $\symb{traffic_class}_{x}$, one can compute the available time for new services $s$ in the set of services $S$ to transmit, $\symb{t_free_k}$ in (\ref{eq:tfree}), where $\symb{t_open}$ is defined within the \ac{gcl} of the \ac{tas} schedule:
\begin{equation}
    \label{eq:tfree}
    \begin{aligned}
    \symb{t_free_k} = \symb{t_open} - \sum_{s\in S} \symb{t_transmission}^{s}
    \end{aligned}
    \raisetag{20pt}
\end{equation}

Considering a tight scheduling approach, if $\symb{t_free} \ge \symb{t_needed}$, the traffic class is capable of handling the new service with minimum effort.
By opposition, if $\symb{t_free} < \symb{t_needed}$, the traffic class open time will necessarily have to be increased.

For the assessment of the requirement compliance, the effort of configuring \ac{tas} for a node can be retrieved as $\symbrank{requirement_cap_i_unnormalized}$, which is then normalized as the output of $\symbrank{requirement_cap}_i()$ below:
{\setlength{\jot}{0.8em}
\begin{align}
\begin{split}
    \symbrank{requirement_cap_i_unnormalized}  = \frac{\symb{t_needed}}{\symb{t_free}}\quad\text{ and }\quad\symbrank{requirement_cap}_i()  = \left\{ \begin{array}{ll}
    	0.5 + \symbrank{requirement_cap_i_unnormalized} / 2 & \text{if } \symb{t_free} \ge \symb{t_needed}\\
    	(\symbrank{requirement_cap_i_unnormalized} - 1) / 2 & \text{if } \symb{t_free} < \symb{t_needed}\\
    \end{array} \right.
    \end{split} \nonumber
\end{align}}

For example, we consider a scenario where a node with a network interface of $\symb{bandwidth}=\qty{1000}{\mega\bit\per\second}$ is currently configured with the following \ac{tas} schedule depicted in figure~\ref{fig:tas-before}. When adding a new service with $\symb{datasize}=\qty{5}{\mega\bit}$ and considering $\symb{t_guard} = \symb{t_transmission} \times 0.01$ (10\% of the transmission time\footnote{This is a value to simplify the example. The study of the appropriated value of $\symb{t_guard}$ is beyond the scope of this work.}):
{\setlength{\jot}{0.6em}
\begin{align}
\begin{split}
    \symb{t_transmission} = \frac{5}{1000} = \qty{5}{\milli\s}\;\text{ and }\; \symb{t_needed} = 5 + 5 \times 0.1 = \qty{5.5}{\milli\s}
\end{split} \nonumber
\end{align}}

\begin{figure}[t]
    \centering
    \includegraphics[width=\columnwidth]{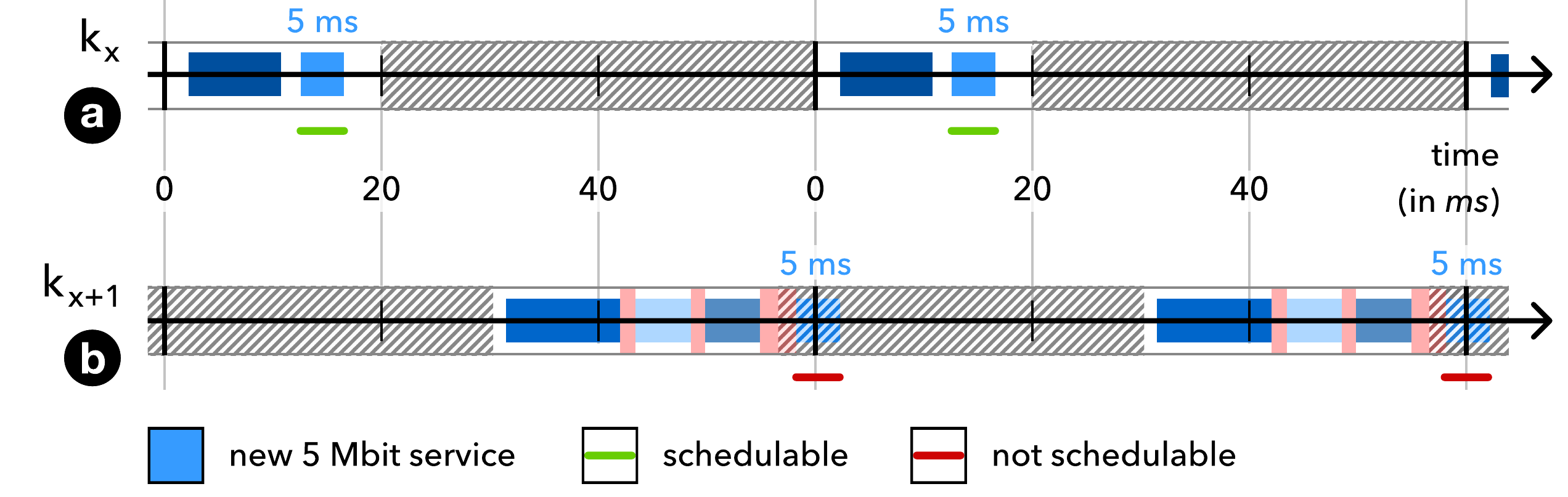}
    %\missingfigure{schedule after inserting on traffic class X and X+1. include new values of T open, Tfree and service transmission times}
    \caption{Impact of new service in class $\symb{traffic_class}_x$ \protect\bcircle{a} or $\symb{traffic_class}_{x+1}$ \protect\bcircle{b}.}
    \label{fig:tas-after}	
\end{figure}

The schedule will be as depicted in figure~\ref{fig:tas-after}, with the values of $\symb{t_free_k}$ and $\symb{t_free_k_plus}$:
\begin{align}
\begin{split}
    \symb{t_free_k} &= 20 - ( 7 ) = \qty{13}{\milli\s}\\
    \symb{t_free_k_plus} &= 30 - ( 13 + 1.3 + 6 + 0.6 + 6 ) = \qty{4.4}{\milli\s}
\end{split} \nonumber
\end{align}

If a scheduling algorithm chooses traffic class $\symb{traffic_class}_{x}$ \bcircle{a} for the new service\footnote{This choice is outside the scope of this work.}, since $\symb{t_free_k} \ge \symb{t_needed}$, the traffic class is capable of handling the new service with minimum effort (a tight scheduling approach). In this scenario, the assessment of the effort of configuring \ac{tas} for the node will be:
\begin{align}
\begin{split}
\symbrank{requirement_cap_i_unnormalized} = \frac{5.5}{13} \approx 0.423\quad\text{ and }\quad\symbrank{requirement_cap}_i() \approx 0.5 + 0.423 / 2 \approx 0.711
\end{split} \nonumber
\end{align}

By opposition, if traffic class $\symb{traffic_class}_{x+1}$ is chosen \bcircle{b}, since $\symb{t_free_k_plus}<\symb{t_needed}$, the traffic class open time will have to be increased and the assessment of the effort of configuring \ac{tas} for the node will be:
\begin{align}
\begin{split}
\symbrank{requirement_cap_i_unnormalized} = \frac{5.5}{4.4} = 1.25 \quad\text{ and }\quad\symbrank{requirement_cap}_i() = (1.25 - 1) / 2 = 0.125 
\end{split} \nonumber
\end{align}

Thus, for the node being assessed, the suitability value \symbrank{bid} will take into account the value of $\symbrank{requirement_cap}_i()$ with the weight as described in section~\ref{sec:design}. Relatively to bare-metal resources assessment, $\symbrank{bid_node_resources}$, the suitability value must reflect the capability of running \ac{tas} within the assessing node. 

This section served as a demonstration of the implementation requirements to add a new resource type in the list of considered types in our proposed resource estimation algorithm. By implementing both criteria of bare-metal resources (which reflects the absolute minimal requirement of a resource) and of current resources (which directly assesses the current availability of a resource), our proposal is able to accept new resource types for assessment.
\section{Conclusions and future work}
\label{sec:conclusions}

This paper proposed a mechanism to evaluate the capability of nodes for accepting an admission resource request list with a graded and normalized score, allowing for full comparability between nodes running the same algorithm in a resource management protocol. Unlike others, the present algorithm incorporates elements concerning the network's current status while assessing admission requests even if no network resources are specified, and allows for new types of resource requirements by overriding the assessment methods for bare-metal resources and evaluating computational, network, or time-related resources relative to the current state.

This proposed algorithm could enhance protocols such as \ac{tsn}'s \ac{rap} mentioned in related work, by accommodating a wider set of requirements for stream characterization, provided they comply with the specified criteria for requirement assessment. Its flexibility allows it to be deployed within other protocols and mechanisms that requires admission control for resource management. The implementation of this algorithm into protocols and mechanisms for resource management allow them to be expandable in terms of new strategies they achieve for resource estimation, and in the expansible possibilities for resource types. Both these features are novel to these resource management mechanisms, as mentioned in section~\ref{sec:related-work}.

In our near future, this proposed solution is to be included in the development of a distributed resource allocation protocol called Rank, that requires this algorithm to facilitate admission requests that consider computing, network, and time-related resources for negotiation and reservation. Moreover, this proposal may not only be used in distributed mechanisms, but also within the context of centralized and orchestrated topologies.

%This paper introduced a set of criteria to allow a computing node to assess its capabilities of accepting an admission resource request list of requirements with a grade. Such a grade, being normalized, can be fully compared between other nodes running the same algorithm within the context of a distributed resource allocation protocol.

%Relatively to others, the present algorithm has both elements concerning the network current status, as well as allow new types of resource requirements to be considered as they will only have to comply by overriding how to assess its bare-metal resource assessment, and how to evaluate its dependable computational, network, or time-related resource, relative to its current state. 

%This proposed algorithm could be an advancement on the protocols stated in the presented related work such as in \ac{rap}, allowing a wider set of requirements to be considered for a stream characterization, as long as they are compliant with the stated criteria for requirement assessment. In future work we intent to develop a distributed resource allocation protocol using this algorithm that would allow admission requests to consider computing-, network-, and time-related resources for negotiation and reservation.

\Urlmuskip=0mu plus 1mu\relax
\bibliographystyle{elsarticle-num} 
\bibliography{vars/bibliography}

\end{document}